\documentclass[10pt, journal, letter, twoside]{IEEEtran}
\usepackage{soul}
\usepackage{multirow}
\usepackage{booktabs} 
\usepackage{ragged2e}
\usepackage[noend]{algpseudocode}     
\usepackage{algorithm}            
\usepackage{threeparttable}
\usepackage{comment}
\usepackage{graphicx}
\usepackage{amsmath}

\usepackage[table,dvipsnames,svgnames]{xcolor}
\usepackage[caption=false,font=normalsize,labelfont=sf,textfont=sf]{subfig}

\usepackage[hidelinks]{hyperref}
\usepackage{xspace}

\algrenewcommand\algorithmicforall{\textbf{foreach}}

\newcommand{\smallerspace}{\vspace{-1em}}
\newcommand{\littlesmallerspace}{\vspace{-0.5em}}
\renewcommand{\smallerspace}{}
\renewcommand{\littlesmallerspace}{}

\usepackage{setspace}

\newcommand{\jk}[1]{{\color{blue}JK: {#1}}}
\newcommand{\fz}[1]{{\color{red}FZ: {#1}}}
\newcommand{\la}[1]{{\color{blue}LA: {#1}}}
\newcommand{\qj}[1]{{\color{red}QJ: {#1}}}
\renewcommand{\jk}[1]{}
\renewcommand{\fz}[1]{}
\renewcommand{\la}[1]{}
\renewcommand{\qj}[1]{}

\newcommand{\TM}{\textit{TroMUX}}
\renewcommand{\TM}{\textit{TroLLoc}\xspace}
\newcommand{\TMMUX}{\textit{TroLLoc-MUX}\xspace}
\newcommand{\TMXOR}{\textit{TroLLoc-XOR}\xspace}

\newcommand{\ML}{\textit{MuxLink}\xspace}
\newcommand{\SC}{\textit{SCOPE}\xspace}
\newcommand{\RS}{\textit{Resynthesis}\xspace}
\newcommand{\RSC}{\textit{Resynthesis + SCOPE}\xspace}
\newcommand{\OM}{\textit{OMLA}\xspace}

\newcommand{\sym}[1]{\parbox[c]{1em}{\includegraphics[height=1em]{incl/figures/sym/sym_#1.png}}}
\newcommand{\symwide}[1]{\parbox[c]{1.5em}{\includegraphics[height=1em]{incl/figures/sym/sym_#1.png}}}

\begin{document}

\newcommand{\thetitle}{\TM: Logic Locking and Layout Hardening for IC~Security Closure
		against Hardware Trojans}

\title{\thetitle}

\author{Fangzhou Wang, Qijing Wang, Lilas Alrahis, Bangqi Fu, Shui Jiang, Xiaopeng Zhang,\\
	Ozgur Sinanoglu, Tsung-Yi Ho, Evangeline F.Y. Young, Johann Knechtel
\thanks{Fangzhou Wang, Qijing Wang, Bangqi Fu, Shui Jiang, Xiaopeng Zhang, Tsung-Yi Ho, and Evangeline F.Y. Young are with the Department of Computer Science and
Engineering, Chinese University of Hong Kong, Hong Kong (e-mail: fzwang@cse.cuhk.edu.hk).}
\thanks{Lilas Alrahis, Ozgur Sinanoglu, Johann Knechtel are with the New York University Abu Dhabi (e-mail: johann@nyu.edu).}%
\thanks{A preliminary version of this work has been presented at the ACM International Symposium on Physical Design (ISPD) in 2023~\cite{Wang23}.}
}

\IEEEtitleabstractindextext{

\begin{abstract}
    Due to cost benefits, supply chains of integrated circuits (ICs) are largely outsourced nowadays.
    However, passing ICs through various third-party providers gives rise to many security threats, like piracy of IC intellectual
    property or insertion of hardware Trojans, i.e., malicious circuit modifications.
    
    In this work, we proactively and systematically protect the physical layouts
    of ICs
    against post-design insertion of
    Trojans.
    Toward that end, we propose \TM, a novel scheme for IC security closure that employs, for the first time, logic locking and layout hardening in unison.
    \TM is fully integrated into a commercial-grade design flow, and
    \TM is shown to be effective, efficient, and robust.
    Our work provides in-depth layout and security analysis considering the challenging benchmarks of the ISPD'22/23 contests for security closure.
    We show that \TM successfully renders layouts resilient, with reasonable overheads, against
    (i)~general prospects for Trojan insertion as in the ISPD'22 contest,
    (ii)~actual Trojan insertion as in the ISPD'23 contest, and
    (iii)~potential second-order attacks where
    adversaries would first (i.e., before Trojan insertion) try to bypass the locking defense, e.g.,
    using advanced machine learning attacks.
    Finally, we release all our artifacts for independent verification~\cite{release}.
\end{abstract}

	\begin{IEEEkeywords}
		Integrated Circuits;
		Hardware Security;
		Trojans;
		Logic Locking;
		Physical Design;
		Security Closure.
	\end{IEEEkeywords}
}

\markboth{IEEE Transactions on Information Forensics and Security}
{Wang \MakeLowercase{\textit{et al.\ }}: \thetitle}

\maketitle

\IEEEdisplaynontitleabstractindextext
\IEEEpeerreviewmaketitle

\section{Introduction}
\label{sec:intro}

\IEEEPARstart{I}{ntegrated}
circuits (ICs) of varying scale and complexity are at the heart of all our modern-day information systems.
An ever-growing body of security threats can compromise information systems in general
and ICs in particular~\cite{rostami14,wu21,knechtel21_Sec_Emerg_ISPD}.
For the design, manufacturing, and deployment of ICs,
there are numerous companies and partners involved within complex and world-wide supply chains---ICs run
through many hands, where some of those may be acting with malicious intent.
Furthermore, once ICs
are deployed in the field,
 adversaries can employ powerful hands-on measures, targeting directly at the hardware and software at runtime.

The resulting cybersecurity challenges are manifold, suggesting the need for holistic and streamlined defense efforts, all the way
from software applications down to the IC hardware.
Toward the latter end, \textit{electronic design automation (EDA)} algorithms and tools will play an important
role~\cite{knechtel20_Sec_EDA_DATE, sigl11, ravi19, wu21}.
However, state-of-the-art (SOTA) commercial EDA tools do not account yet for
such security challenges.

\textbf{Security closure} is an emerging paradigm that
seeks to address this shortcoming of EDA tools. The idea is to proactively harden the layouts of ICs,
during physical synthesis, against various threats that are executed post design-time~\cite{knechtel22_SCPL_ISPD,knechtel21_SCPL_ICCAD,eslami23}.
Such efforts are important for multiple reasons. For example,
threats like hardware Trojans or side-channel
attacks are explicitly exploiting vulnerabilities of the IC layouts, and any such threats that are not mitigated during design-time will be very difficult to address later on---unlike software, ICs are not easily patchable.
See Sec.~\ref{sec:bg:SCPL} for more background.

In this work, we focus on the threat of \textbf{hardware Trojans}, i.e., malicious circuit modifications
that can be introduced by different adversaries throughout the IC supply chain.
By design, Trojans are only minor in extent but severe in
fallout~\cite{xiao16,dong20,yang16_a2}.
Many different types of countermeasures have been proposed over the years.
However, most prior art falls short in terms of effectiveness, efficiency, and/or robustness.
See Sec.~\ref{sec:bg:Trojans} for more background on Trojans and
Sec.~\ref{sec:prior} for a discussion of relevant prior countermeasures, respectively.

Besides, \textbf{logic locking}, or locking for short, means to
incorporate so-called key-gate structures
that are controlled by secret key-bits~\cite{YRS20}.
Without
knowing the related key-bits required to unlock those key-gate structures, attackers cannot fully understand the structure and functionality of the
design at hand.
While locking is mainly known for protection of chip-design intellectual property (IP),
it can also serve for Trojan defense.
See Sec.~\ref{sec:bg:LL} for more background on locking.

The \textbf{objective of this work} is security closure against hardware Trojans.
That is,
we aim for \textit{proactive, pre-silicon Trojan prevention, by carefully and systematically hardening the physical
layout of ICs against post-design Trojan insertion}.

Based on a thorough review of prior art and its limitations,
	we derive important \textbf{research challenges} as follows.
	Note that various concepts indicated on here, like second-order attacks or open placement sites, are all explained in detail in Sec.~\ref{sec:bg}.
\begin{itemize}

\item \textit{Robustness -- Any Trojan defense
must remain robust in place.}
That is, a foundry-based adversary, being fully aware that some defense is put in
place, would naturally want to first circumvent that defense
(i.e., stealthily remove, override, or otherwise render useless)
before inserting their Trojans.
Such {second-order attacks represent a key challenge where
prior art
	falls short.}

\item \textit{Effectiveness -- Any defense should protect against various Trojans.}
This is especially true for proactive, pre-silicon defenses which have
only ``one shot'' at design-time.
More specifically, {designs should be protected as a whole}, i.e., in terms of (i)~layout resources
exploitable for Trojan insertion in general (i.e., open placement sites, free routing tracks, available timing
	slacks) and (ii)~the netlist structure and functionality, which is relevant
to hinder Trojans targeting on specific assets.
Furthermore, defenses should be evaluated against actual layout-level Trojan insertion, not only against vulnerability estimates.

\item \textit{Efficiency -- Any defense should incur limited overheads.}
Taking control over the underlying trade-offs for resilience against Trojans versus design overheads {requires} two parts:
{(i)~{metrics for security and overheads}, and (ii)~some integrated, {secure-by-design EDA methodology},}
utilizing those metrics and some user guidance.

\end{itemize}

\begin{figure}[tb]
\centering
\includegraphics[width=.90\columnwidth]{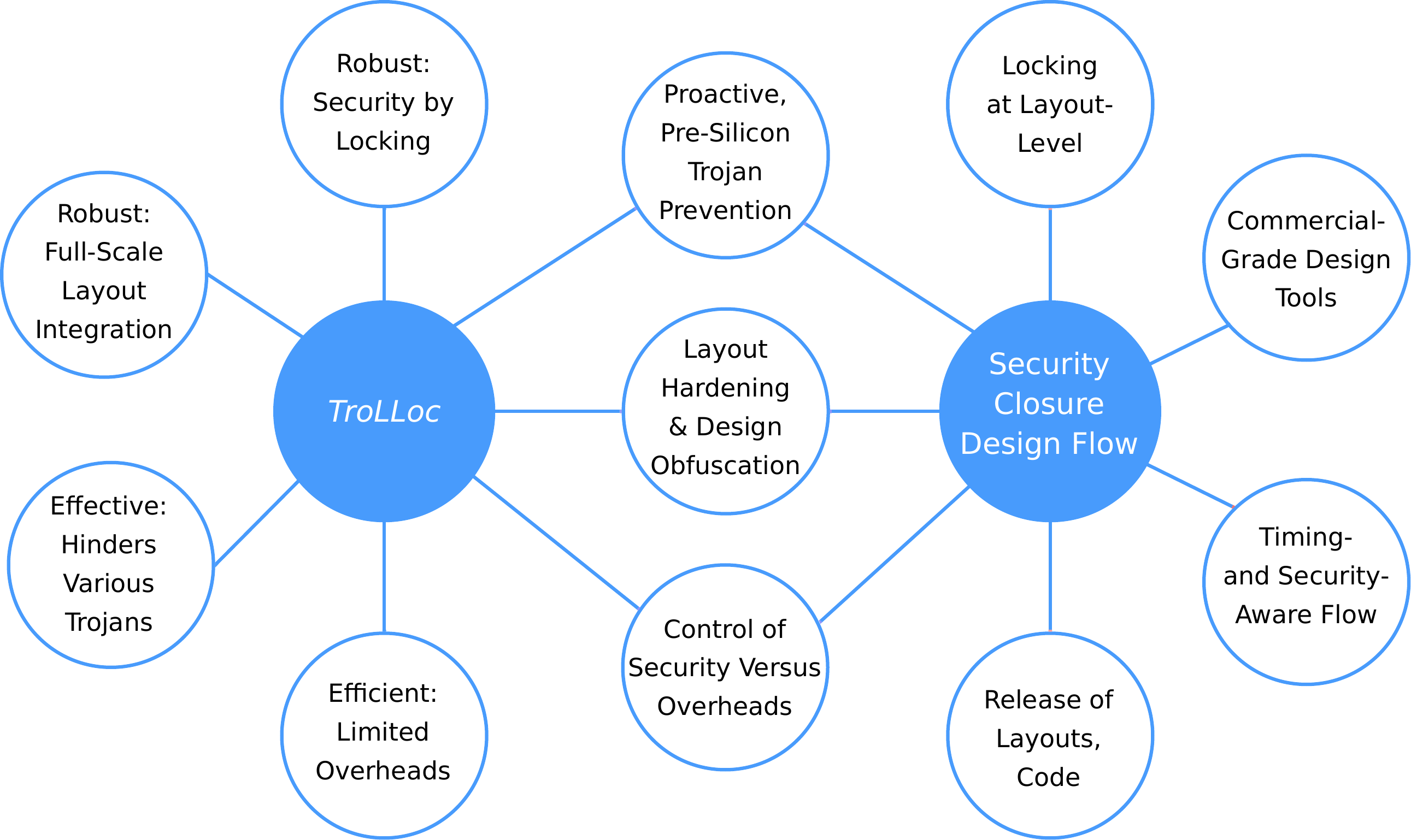}
\littlesmallerspace
\caption{High-level overview of work. The two main contributions are the locking scheme, \TM, and the IC {security-closure} flow. Both are carefully orchestrated together to achieve pre-silicon Trojan prevention.
}
\label{fig:overview}
\smallerspace
\end{figure}

To address all these challenges,
our work makes the following
\textbf{contributions}. A high-level representation of our contributions is also given in Fig.~\ref{fig:overview}.
\begin{enumerate}

\item Layout-level logic locking
--
	We propose a novel locking scheme, called \TM, devised to hinder post-design Trojan
	insertion.
	Together with the IC security-closure flow, \TM meets important challenges for Trojan defenses:
		it is robust, effective, and efficient. See further below for an explanation of these achievements.

\item IC security-closure flow --
	We devise an EDA flow for design-time security closure against post-design Trojan insertion.
	The flow serves to implement the proposed \TM scheme and is fully integrated
	into a commercial-grade physical-synthesis setup.

\item Layout and security analysis --
	We conduct a set of thorough case studies that show the promises of the proposed scheme.
	Toward that end, we are using the real-world benchmark suites from the recent ISPD'22/23 contests on security closure~\cite{knechtel22_SCPL_ISPD,eslami23}.
	We are employing a set of SOTA attacks on locking,
 namely \ML~\cite{alrahis22_ML}, \OM~\cite{alrahis21_OMLA}, \RS~\cite{almeida23}, and \SC~\cite{alaql21},
 and we also conduct
	actual Trojan insertion as in \cite{eslami23}, both to demonstrate the resilience of the proposed scheme against advanced Trojan attacks.

\item Release --
	We release our secured layouts in~\cite{release}, to
	enable independent verification of our work.
	We will also release our method in full, post peer-review.

\end{enumerate}

	Our work on proactive, pre-silicon Trojan prevention is (i)~robust, (ii)~effective, and (iii)~efficient.
	Regarding (i),
	we show that advanced adversaries employing (a)~second-order attacks using SOTA machine learning (ML)-based attacks on locking or (b)~actual Trojan insertion are both hindered.
	This is a first time in the literature.
	Regarding (ii), also for the first time, we harden IC layouts against Trojan insertion in general and further protect security-sensitive assets in particular.
	Regarding (iii), again for the first time, we employ both locking and layout hardening in unison, along with commercial-grade tooling for full control on the security-vs-overhead trade-offs in a real-world IC
	design and manufacturing setting.

\section{Background and Motivation}
\label{sec:bg}

\subsection{Trojans}
\label{sec:bg:Trojans}

\subsubsection{Real-World Relevance}

ICs should provide a root-of-trust platform on which sensitive data operation can rely on.
If that premise is broken, however, the confidentiality, integrity, and availability of any application is entirely at risk.
For consumer electronics, Trojans that aim to compromise cryptographic circuitry, which exists in modern processors more many years by now, represent such real-world risks~\cite{perez22}. 
For other domains like military or avionics, the stakes are even higher. Potential Trojan threats on military-grade ICs are discussed in~\cite{adee08,Skorobogatov12}.
Remarkable case studies for Trojans in real ICs are described in~\cite{muehlberghuber13, yang16_a2, ghandali20, perez22, Puschner23}.

\subsubsection{Working Principles}

Trojans are malicious hardware modifications~\cite{xiao16,dong20}.
This concept is quite diverse, covering malicious modifications that may:
(i)~leak information from an IC, reduce its performance, or disrupt its working altogether;
(ii)~be always on, triggered internally, or triggered externally;
(iii)~be introduced through untrustworthy third-party IP, adversarial designers, mask generation,
or manufacturing of ICs;
etc.
Besides, most Trojans require some additional layout-level resources like \textit{open placement sites}, \textit{free routing
tracks}, and/or \textit{timing slacks}~\cite{knechtel22_SCPL_ISPD,eslami23,trippel20,salmani16}.\footnote{%
	Modern IC layouts utilize \textit{standard-cell libraries}, which are provided by foundries and their suppliers, along with the
		technology-dependent design rules in \textit{technology libraries}.
	       As indicated, the cells in those libraries are standardized, namely in their geometries,
       such that cells are relatively easy to place into regularly sized and arranged placement rows of an IC.
	Placement sites are bins of fixed width within such rows. Thus,
       open placement sites are bins that are not occupied yet by some standard cell.
	Further, an IC's interconnect comprises a stack of different metal layers; the latter are arranged into regular horizontal and vertical routing tracks. Thus, free routing tracks are stripes within
	these metal layers that are not utilized yet by some routing shape.
	Timing slacks can be thought of as performance margins. For an example in simple terms, assume that some part of an IC design, which does exhibit timing slacks, may become slower,
	e.g., due to insertion of additional Trojan logic. Then, the performance constraints of the overall design can still be met as long as the introduced slow-down is less than the available slacks.}

Most if not all Trojans comprise a \textit{trigger} and a \textit{payload}; the trigger
activates the payload on some attack condition, and the payload serves to perform an actual attack.
Triggers are often based on \textit{low-controllability nets (LCNs)}\footnote{\textit{Controllability} describes whether an end-user can establish specific signal assignments for some internal nets through the IC's
	primary inputs. Thus, LCNs are nets where end-users are challenged, across a wide range of input patterns, to achieve specific signal assignments and toggling activity.}---to hinder their detection during
	testing, rendering them more stealthy---whereas
payloads are targeting on sensitive assets like registers holding some key.
Note that the trigger and payload components are implemented individually but are working together in tandem.
For example, once some LCN is toggling, i.e., only for some rare and carefully crafted condition, the trigger component initiates the actual attack, e.g., the content of some sensitive registers is maliciously forwarded
to non-privileged circuit parts or to primary outputs, or even more stealthily leaked through side-channels~\cite{perez22}.

\subsubsection{Threat Modeling}
\label{sec:bg:Trojans:TM}

We assume post-design insertion of additive Trojans by adversaries residing within foundries.
Among different scenarios (e.g., see~\cite{xiao16,dong20} for some taxonomies),
this one is outstanding as powerful and versatile---related Trojans are not detectable at all during design-time and are also not limited to, e.g., denial-of-service attacks through removal or modification of existing logic, but can embed any
malicious functionality at will.

More details on the threat modeling for our work are given in Sec.~\ref{sec:tm}.
Besides, an important novel concept for threat modeling in our work is introduced next.
While similar ideas may have been used in the literature before, we are the first to explicitly formulated this concept as such.

We differentiate between \textit{first-order} versus \textit{second-order} Trojan attacks. For the former, adversaries would seek to directly insert their Trojans into the IC layout under attack.
For the latter, adversaries would first revise the layout to regain layout resources, to aid subsequent Trojan insertion. Both attacks can be supported by means of \textit{engineering change order (ECO)}
techniques~\cite{perez22,AlexTRJ,wei24}.\footnote{%
	In general, ECO techniques are an industry-wide standard for
		incremental changes of IC layouts, e.g., to revise the placement and routing of a small module within a larger IC design.
		Their (mis-)use for Trojan insertion has been successfully demonstrated only recently~\cite{perez22,AlexTRJ,eslami23,wei24}.}
Besides, in case some layout defenses are put in place, conducting second-order attacks may require adversaries to bypass those as well. Depending on the nature of the
defense, it may suffice to again utilize ECO techniques, e.g., to remove \textit{fillers} and/or \textit{spares},\footnote{%
	Spares are redundant standard cells; they are placed early on but not fully connected yet.
	Such spares are helpful for last-minute ECO procedures, as their use would likely require only routing-level changes.
	Fillers are various non-functional cells that serve different purposes, e.g., acting as decoupling capacitors (decaps) to stabilize the IC's internal power supply network.
	Although spares and fillers can be used for pre-silicon prevention, by simply filling up all open placement sites with them,
	it is important to note that those components are easy to exploit for Trojan insertion as most of them can be readily removed without disrupting the core functionality of an IC~\cite{eslami23,Puschner23}.
} to rearrange the cell placement, etc.
More advanced efforts may also be required; e.g., dedicated attacks would be needed to decipher and remove locking structures.

\subsubsection{Countermeasures and Their Challenges}
\label{sec:bg:prior}

Trojan countermeasures can be classified into:
\begin{enumerate}
\item Proactive, pre-silicon prevention 
	(e.g.,
	 \cite{dupuis14,marcelli17,samimi16,sisejkovic19,xiao14,ba15,ba16,knechtel21_SCPL_ICCAD,hosseintalaee17,imeson13,shi18,trippel23,Guo23,Hsu23,Eslami23a,wei23,Eslami24});
\item Reactive, post-silicon detection or monitoring 
	(e.g., \cite{guimaraes17,hou18,vijayan20,kim11_trojan,basak17,wu16,wahby16,Agrawal07,Eslami24});
\item Proactive detection, that is
	(a)~pre-silicon inspection/verification
	(e.g., \cite{dong20,guo19_QIFV,fern17,chen18,Lashen23}),
	(b)~post-silicon testing (e.g., \cite{narasimhan13,dong20,deng20}), or (c)~post-silicon layout inspection (e.g., \cite{sugawara14, vashistha18, Puschner23}).
\end{enumerate}
There are various challenges for each of this class as follows.
First, many post-silicon schemes rely on a ``golden IC,'' i.e., a Trojan-free reference IC~\cite{Agrawal07}. In threat models where the foundry is untrusted, this requirement is impractical.
Second, both post-silicon monitoring and pre-silicon prevention require dedicated hardware features---if not secured properly,
the related circuitry itself may be circumvented by adversaries during second-order attacks.
Third, by construction, pre-silicon inspection/verification cannot tackle any Trojans inserted by the foundry---such Trojans do not exist yet at design-time.
Fourth,
post-silicon testing is challenging for multiple reasons:
	(i)~Regular testing procedures
	may not be helpful to determine some rarely triggered functional or parametric divergences introduced by Trojans.
	(ii)~Trojans can be realized with largely varying logic structures and functional behaviour, making an identification of ``common patterns'' difficult~\cite{gohil22_ATT}.
	While recent ML works like~\cite{kento17,Lashen23} show promise toward that end, they also cannot achieve perfect accuracy.
	(iii)~Process variations as well as measurement noises make the observation of parametric effects introduced by Trojans more and more challenging for advanced IC technologies~\cite{jacob14}.

Given all these challenges, we argue that {a \textit{robust} scheme for proactive prevention is
essential to try to hinder Trojans early on and as best as possible.}
We discuss the relevant prior art
in more detail in Sec.~\ref{sec:prior:locking} and Sec.~\ref{sec:prior:layout}.

\subsection{Security Closure}
\label{sec:bg:SCPL}
	As indicated,
	security closure
	seeks to proactively harden the physical layouts of ICs
	at design-time against various threats that are executed post design-time~\cite{knechtel21_SCPL_ICCAD, knechtel22_SCPL_ISPD,eslami23}.
	Security closure against Trojans means
to control physical synthesis such that insertion of Trojan
components
is prevented altogether or at least
becomes impractical, while managing the impact on design quality~\cite{knechtel21_SCPL_ICCAD,knechtel22_SCPL_ISPD,eslami23,Hsu23,Guo23,wei23}.
For example, aggressively dense layouts would leave only few open placement sites
and few routing resources
exploitable for Trojan insertion~\cite{eslami23}.
However,
while such layouts are already challenging by themselves, in particular for managing
\textit{design rule checks (DRCs)},\footnote{%
DRCs describe technology rules for manufacturability of ICs, in terms of cell placement (e.g., considering the so-called \textit{N-well continuity} which is essential for IC semiconductors) and in terms of routing shapes
	(considering minimum areas, pitches, et cetera).
Thus, any violation for DRCs can disqualify the whole IC from manufacturing.
For modern ICs and their advanced technologies underlying, DRCs are becoming more and more complex~\cite{baek22}.}
such arguably naive approach may not be good enough for security closure.
This is because, for one, advanced Trojans like \textit{A2}~\cite{yang16_a2} may require only 20 placement sites~\cite{trippel20};
that few sites will remain even in any aggressively dense layouts, especially once filler and/or spare cells are considered as exploitable open sites as well.
For another, recall that {second-order attacks} can regain layout resources before Trojan insertion, thereby rendering such implicit protection postulated by dense layouts ineffective in general.

Overall, we argue that {efforts for security closure against Trojans need to address the challenges outlined in
	{Sec.~\ref{sec:intro}}}.

\begin{table*}[tb]
	\footnotesize
\setlength\tabcolsep{6.0pt}
	\caption{High-Level Comparison with Prior Art for Pre-Silicon Trojan Prevention
	}
	\label{tab:hl_comp}
	\littlesmallerspace
	\littlesmallerspace
	\littlesmallerspace
	\center
	\begin{tabular}{cc|ccccc}
		\toprule
		\multirow{2}{*}{\textbf{Works}} & \multirow{2}{*}{\textbf{Approach}} & \multirow{2}{*}{\textbf{Robust}} &
		{\textbf{Effective Protection of:
			}} & \multirow{2}{*}{\textbf{Efficient}} & \textbf{Tested against} & \textbf{Artifacts} \\
		& & & \textbf{Functionality / Layout} & & \textbf{Actual Trojans} & \textbf{Available} \\
		\midrule
		\cite{dupuis14} & Locking & \sym{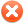}~\cite{dupuis14,samimi16} & \sym{no}$^{a}$ / \sym{no}~~~ & \symwide{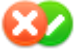} & \sym{no} & \sym{no} \\
\rowcolor{Gainsboro}
		\cite{samimi16,marcelli17} & Locking & \symwide{_no_}~\cite{alrahis21_OMLA} & \symwide{_no_}$^{a}$ / \sym{no}~~~~ & \symwide{_no_} & \sym{no} & \sym{no} \\
		\cite{sisejkovic19} & Locking & \sym{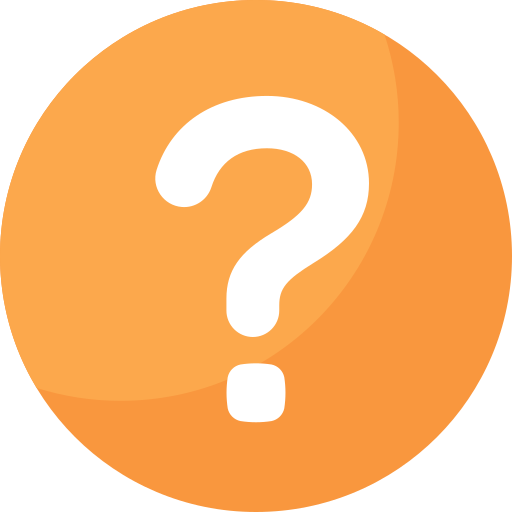}~ & \sym{_}$^{a}$ / \sym{no}~~~ & \symwide{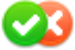} & \sym{no} & \sym{no} \\
\rowcolor{Gainsboro}
		\cite{xiao14,ba15,ba16,Eslami24} & Additional Logic & \symwide{_yes_}~\cite{perez22,AlexTRJ,eslami23} & \sym{no} / \symwide{_yes_}$^{a}$ & \symwide{_yes_} & \sym{no} & \sym{no} \\
		{\cite{hosseintalaee17,knechtel21_SCPL_ICCAD,Guo23,Hsu23,Eslami23a,wei23}} & Layout Hardening & {\symwide{_no_}~\cite{perez22,AlexTRJ,eslami23}} &
		{\sym{no} / \symwide{_no_}$^{a}$} & {\symwide{_yes_}} & \sym{no} & {\symwide{_yes_}} \\
\rowcolor{Gainsboro}
		{\cite{trippel23}} & Routing & {\sym{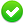}} &
		{\sym{no} / \symwide{_yes_}$^{b}$} & {\sym{yes}} & \symwide{_yes_}~\cite{yang16_a2} & {\sym{no}} \\
		\multirow{2}{*}{\textbf{Ours: \TM}} & \textbf{Locking and} & \multirow{2}{*}{\textbf{\sym{yes}}} &
		\multirow{2}{*}{\textbf{\sym{yes} / \sym{yes}}}~~ & \multirow{2}{*}{\textbf{\sym{yes}}} & \multirow{2}{*}{\textbf{\sym{yes}~\cite{eslami23}}}
		& \multirow{2}{*}{\textbf{\sym{yes}~\cite{release}}} \\
		& \textbf{Layout Hardening} & & & & \\
		\bottomrule
	\end{tabular}\\[1mm]
	\begin{justify}
Symbols: \sym{yes} -- yes, \sym{no} -- no, \sym{_} -- unclear, \symwide{_yes_} -- yes but with some caveats, \symwide{_no_} -- unlikely.
Notes:
	   ${a}$ -- Lack or uncertainty of robustness is expected to undermine the real-world effectiveness.
	$b$ -- Covers only routing resources.
	\end{justify}
\smallerspace
\end{table*}

\subsection{Logic Locking}
\label{sec:bg:LL}

\subsubsection{Working Principles}
As indicated, logic locking or locking for short, means to incorporate so-called key-gate structures that are controlled by a secret key~\cite{YRS20}.
Without the full knowledge of the key, logic locking ensures that the details of the design IP cannot be fully recovered and the IC remains non-functional. 
The key-gate structures are commonly realized using, e.g., X(N)OR gates~\cite{roy10}, AND/OR
gates~\cite{dupuis14}, look-up tables (LUTs)~\cite{8429401}, or multiplexers (MUXes)~\cite{rajendran15}.
After manufacturing, preferably only after testing~\cite{yasin16_test}, the IC is activated: the secret key is loaded into a dedicated, on-chip tamper-proof memory by a trusted party. 

While locking is largely known for protection of design IP~\cite{YRS20}, it can also serve for Trojan defense. In fact, different such schemes have been proposed: AND/OR locking~\cite{dupuis14}, X(N)OR
locking~\cite{samimi16,marcelli17}, and custom locking~\cite{sisejkovic19}.
	We discuss relevant prior art in more detail in Sec.~\ref{sec:prior:locking}.

\subsubsection{Threat Modeling}
Over the years, various attacks against the different implementations of locking have been proposed.
Two different threat models are considered:\footnote{%
For our work, we only have to consider the oracle-less model. This is because we seek to protect against post-design Trojan insertion by the foundry, where oracle ICs are not available yet.
More details on the threat modeling for our work are given in Sec.~\ref{sec:tm}.
}
\begin{itemize}
\item The \textit{oracle-guided} model assumes access to an activated IC (i.e., with the correct key loaded), which can be purchased from the market, yet only after manufacturing.
Using another IC copy, the attacker also has to extract the locked gate-level netlist through
means of reverse engineering. The attacker then aims to decipher the key by analyzing the netlist, identifying meaningful input patterns through analytical means such as Boolean satisfiability (SAT)
	solving~\cite{Subramanyan2015}, applying those to the working IC, and observing the responses to iteratively converge toward the correct key.
\item The \textit{oracle-less} model assumes only access to the locked gate-level netlist.
The attacker aims to decipher the key by analyzing the structure and functionality of the netlist, e.g., by means of ML~\cite{alrahis22_ML}, identifying
meaningful patterns and correlations of key-gates
and correct key-bits.
\end{itemize}

\subsubsection{Advanced Oracle-Less Attacks}
Recently, ML-based attacks like \textit{SAIL}~{\cite{chakraborty18}}, \SC~{{\cite{alaql21}},
\OM~{\cite{alrahis21_OMLA}} and \ML~\cite{alrahis22_ML}, as well as structural attacks like
\RS~\cite{almeida23},
succeeded in identifying various relevant design features in locked circuits and in subsequently predicting key-gate structures, all in an oracle-less setting.
	
For example, using a graph representation of the IC design, \ML~\cite{alrahis22_ML}
learns on the structure of regular, unlocked parts of a design at hand that is specifically locked using MUX key-gates.
After training, \ML seeks to decipher the input connections to MUX key-gates.
More specifically, \ML considers each MUX key-gate's input at a time, connecting it to the output of the MUX---thereby bypassing the key-gate structure---and contrasts the resulting different
outcomes to the trained knowledge to predict the more likely original structure, i.e., the circuitry before locking.

For another example, \RS~\cite{almeida23}
	systematically resynthesizes the locked netlist across a large range of varying synthesis parameters like mapping efforts and timing constraints,
the latter especially for paths related to key-gate structures.
All resulting netlists, which are structurally different yet functionally equivalent, are subsequently passed to other attacks like \SC, upon which majority voting is conducted across all the resulting inferences
for each key-gate structure.

Given their high success rates, such {oracle-less attacks represent a considerable threat for locking. Especially 
for locking-based, proactive Trojan prevention at design-time, such attack capabilities need to be taken in consideration. Otherwise, second-order attacks would remain a practical threat.
 
\section{Prior Art and Their Limitations}
\label{sec:prior}

Next, we discuss the prior art for proactive, pre-silicon Trojan prevention in more detail, including their limitations. We differentiate into locking-based versus layout-based schemes.

A high-level comparison of prior art and ours is also outlined in Table~\ref{tab:hl_comp}.
It is important to note that our work addresses \textit{all} the limitations identified for prior art.

\subsection{Locking-Based Trojan Prevention}
\label{sec:prior:locking}

Dupuis et al.~\cite{dupuis14} lock LCNs using AND/OR key-gates, to hinder insertion of Trojan triggers. They consider timing slacks, varying toggling thresholds, and balanced switching probabilities.
Samimi et al.~\cite{samimi16} follow similar principles as Dupuis et al., but utilize X(N)OR key-gates.
Marcelli et al.~\cite{marcelli17} propose a multi-objective evolutionary algorithm, seeking to
minimize LCNs and maximize the efficacy of X(N)OR locking at the same time.
\v{S}i\v{s}ejkovi\'{c} et al.~\cite{sisejkovic19} secure inter-module control signals against
software-controlled hardware Trojans, using encryption circuitry along with regular locking.
However, they do not specify/limit the type of key-gates.

The above prior art has limitations as follows.
First,	Dupuis et al.~\cite{dupuis14} cannot protect specific assets against Trojan insertion. This is because they
utilize AND/OR key-gates, to tune the controllability of selected nets, but not to obfuscate the design; see also the next point.
Second, none show robustness against SOTA oracle-less attacks on locking.
For example, Dupuis et al.~\cite{dupuis14}
do not employ resynthesis after locking which, however, would be essential to obfuscate the correlation between key-gate structures and key-bit assignments~\cite{YRS20}.
Third,	\v{S}i\v{s}ejkovi\'{c} et al.~\cite{sisejkovic19} can only protect against Trojans targeting on
inter-module control signals, not on any of the modules' internal circuitry. Accordingly, Trojan attacks seeking to, e.g., manipulate certain computation steps, can still succeed.
Fourth, none consider Trojans at the layout level; all are working at netlist level.
This implies that none can conclusively prevent post-design insertion of Trojans in the real world.
Finally, none explicitly prevent Trojan payloads.
\v{S}i\v{s}ejkovi\'{c} et al.~\cite{sisejkovic19}
are locking the design in general, without focus on triggers or payloads, and others are locking
only LCNs to hinder the integration of Trojan triggers.

\subsection{Layout-Based Trojan Prevention}
\label{sec:prior:layout}

Xiao et al.~\cite{xiao14} propose to add built-in self-test components, arguing that tampering of
those structures (e.g., by adversaries trying to regain layout resources for their Trojans) can be detected during
post-silicon testing.
Similarly, Ba et al.~\cite{ba15,ba16} and Eslami et al.~\cite{Eslami24} insert other additional circuitry.
Hossein-Talaee et al.~\cite{hosseintalaee17} redistribute whitespace / open placement sites otherwise exploitable for
Trojan insertion.
Trippel et al.~\cite{trippel23} employ specific routing structures, to prevent routing of Trojan components.
Knechtel et al.~\cite{knechtel21_SCPL_ICCAD}, Guo et al.~\cite{Guo23}, Hsu et al.~\cite{Hsu23}, Wei et al.~\cite{wei23}, and Eslami et al.~\cite{Eslami23a} all seek to harden layouts through physical synthesis.
For example, Knechtel et al.\ propose to locally increase placement density (to hinder insertion of Trojans) and
to locally increase routing density (to hinder routing of Trojans), while
Eslami et al.\ propose a holistic synthesis methodology; they demonstrate their work with an actual IC tape-out as well.

The above prior art has limitations as follows.
First, none can truly protect specific assets against Trojans. In the absence of
locking or other obfuscation schemes, the original design is fully accessible by adversaries.
Second, except for Trippel et al.~\cite{trippel23}, none show robustness against second-order attacks.
For example, for Hossein-Talaee et al.~\cite{hosseintalaee17}, shifting of whitespaces could be trivially reverted by ECO-assisted Trojan insertion~\cite{perez22,AlexTRJ,wei24}.
While the works of \cite{xiao14,ba15,ba16} aim toward such robustness, the use of additional circuitry by itself does not resolve the fact that other parts of the IC design may remain exploitable.
In particular, fillers and spares are utilized in most if not all prior art and, as indicated in Sec.~\ref{sec:bg:Trojans:TM}, such components are easy to exploit.
Third,
for full efficacy, the work by Xiao et al.~\cite{xiao14} requires 100\% test coverage which can be very difficult to achieve for real-world, large-scale designs.
Further, high utilization rates are challenging for their method. Both aspects are limiting the practicability of the scheme.
Fourth, for the works in \cite{xiao14,ba15,ba16}, the number of additionally required primary inputs scales with layout hardening.
This is impractical; pads for primary inputs/outputs (PIs/POs) are large in actual ICs and, if not employed wisely, can considerably increase the chip outline, directly increasing cost for silicon area.
Finally, except for Trippel et al.~\cite{trippel23}, none have evaluated the resilience of their proposed schemes against insertion of actual Trojans. Note that Trippel et al.\ consider only one specific Trojan,
namely \textit{A2}~\cite{yang16_a2}; their evaluation cannot provide insights for any other kinds of Trojans.

\section{Threat Modeling}
\label{sec:tm}

\subsubsection{Trojans}
We base our assumptions on classical threat modeling for post-design insertion of additive Trojans~\cite{xiao16,muehlberghuber13,trippel20,trippel23,perez22}.
	Details are given next.
\begin{enumerate}

\item We assume adversaries reside within foundries, whereas the design process is 
trustworthy.

\item From 1) follows that adversaries have no knowledge of the original design, only of the
physical layout at hand, which is protected through \TM.
Adversaries also have full knowledge of the IC technology.

\item In an advancement over prior art, we do not utilize any spares or fillers in our protected layouts. Thus, naive Trojan attacks exploiting such resources are ruled out.

\item From 1) also follows that we do not account for Trojans introduced by, e.g., malicious third-party IP
modules.
	Orthogonal countermeasures,
like reactive monitoring 
(Sec.~\ref{sec:bg:prior}), may be applied against such threats as well, in conjunction to ours.

\item Trojans are assumed to be implemented using regular standard cells, not by modifying
transistors, etc.

\item We assume Trojans with triggers based on LCNs and payloads targeting
on specific \textit{assets}, i.e., security-critical components like registers holding some key.

\item The adversaries' objective is to insert
some Trojan(s), more specifically some dedicated trigger and payload components,
into the physical layout.
	Toward that end, adversaries may follow first-order attacks, i.e., direct insertion of Trojan components in the layout as is, or second-order attacks, i.e., revise the layout as much as
	necessary to insert the Trojan,
	as long as the layout then still meets DRCs and other final design checks.\footnote{%
	While reverse engineering and/or reactive monitoring may
	help to detect the related layout modifications introduced by second-order attacks---as well as to detect Trojans in general---we do not consider such orthogonal post-silicon techniques as part of our work.
	}

\item Related to 7),
		recall that attackers cannot exploit fillers and/or spares, as our \TM-protected layouts do not contain such components.

\end{enumerate}

\subsubsection{Locking}
Recall that {we employ locking exclusively to hinder Trojan insertion.} 
Thus, we apply oracle-less threat modeling~\cite{alrahis22_ML,chakraborty18,alaql21,alrahis21_OMLA,almeida23}, in consistency with the above threat modeling for Trojans, as follows.
\begin{enumerate}

\item Adversaries are assumed to reside within foundries. Thus, {only oracle-less attacks
are applicable.}

\item From 1) follows that adversaries have no knowledge of the original design, only of the
physical layout at hand, which is protected through \TM.
Adversaries also have full knowledge of the IC technology.

\item Following Kerckhoffs' principle, all implementation details for \TM are known to the adversaries;
only the secret key-bits remain undisclosed.

\item The adversaries' objective is to circumvent the locking scheme, to (a)~understand the true functionality of the
design, which is required for attacking specific assets, and (b)~to regain layout resources, which is required for Trojan insertion in general.

\end{enumerate}

\section{Methodology}
\label{sec:method}

\begin{figure*}[tb]
	\centering
	\includegraphics[width=0.73\textwidth]{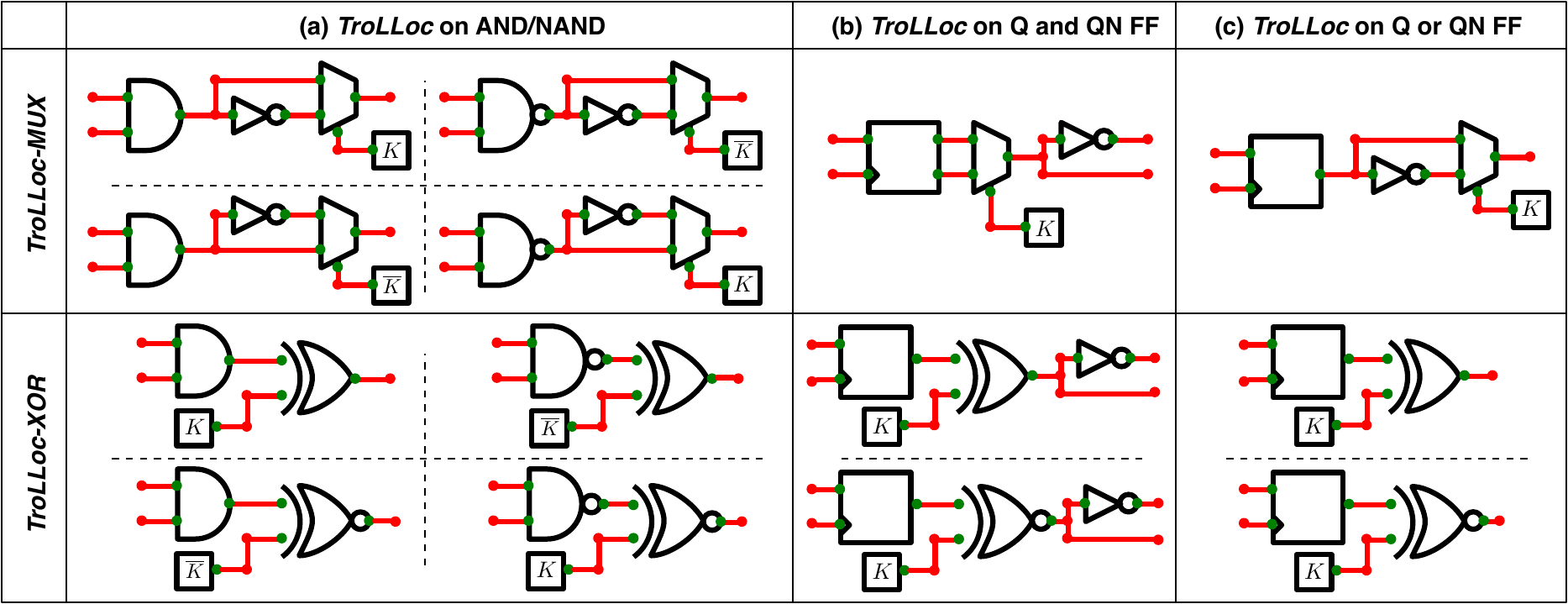}
\littlesmallerspace
	\caption{Design of \TM instances. For \TMMUX or \TMXOR, respectively, the key-bit is connected to the MUX select line or an XOR/XNOR input.
	(a) Locking of simple AND/NAND gates. Note that other simple gates are locked similarly, which is not illustrated here.
	When $K=0$, all \TM instances operate as AND gate; when $K=1$, they function as NAND gate.
	Therefore, given any such \TM instance, its functionality will remain unknown without the correct key-bit.
	(b, c) Locking of different types of FFs.
	For (b), the key-bit dictates which output signal holds Q and which QN;
	for (c), the key-bit dictates whether Q or QN is put out.
	For \TMXOR, we randomly pick only one of the FF's outputs and randomly pick XOR or XNOR as key-gates.
	}
	\label{fig:trolloc_merged}
\smallerspace
\end{figure*}

Recall that we seek to devise a scheme for proactive, pre-silicon Trojan prevention as an addition/alternative to only reactive, post-silicon Trojan detection.
Also recall that related prior art all falls short in terms of robustness, effectiveness, and/or efficiency.
Our approach---which can be summarized as locking and layout hardening applied carefully in unison---aims to hinder both first-/second-order
attacks of post-design Trojan insertion.
For locking, we devise a scheme which is (i)~resilient against various SOTA attacks
and (ii)~directly integrated at layout-level.
For layout hardening, unlike prior art, we neither
employ trivial filler/spare cells nor any other vulnerable circuitry but only locking instances.
To protect IC layouts holistically, our methodology carefully embeds, in a security-aware manner, as many locking
instances during physical synthesis as practically possible, i.e., while keeping the design quality well under control.

Next, we propose the locking scheme, \TM. Thereafter, we devise its integration into a security-and-design-aware synthesis flow based on commercial-grade IC tooling.

\subsection{\TM}

\subsubsection*{Locking Approach}

First, note that both MUX and X(N)OR gates are reasonable choices for key-gates in locking schemes. In fact, both types are used extensively in prior art, with specific attacks being tailored for each type (Sec.~\ref{sec:bg:LL}).
	Thus, we consider both types for implementation as well as all the experimental investigation.
For brevity, we will refer to ``\TM implemented with MUXes as key-gates'' as \TMMUX and ``\TM implemented with X(N)ORs as key-gates'' as \TMXOR in the remainder of this paper.

Unlike prior art for locking-based Trojan prevention (Sec.~\ref{sec:prior:locking}),
\TM utilizes key-gates in a similar way as done in SOTA locking schemes, with the specific intent to
avoid information leakage arising from the key-gates' structure and connectivity.
\TM employs simple but resilient key-gate structures with localized connectivity, described next.

\subsubsection*{
	Learning-Resilient Key-Gate Structures}
\TM ensures that key-bits can be fully randomized, without inducing correlations to the locked gate's
true functionality. Thus, \TM shall be resilient against ML-based attacks by construction.
To do so, \TM instances render any locked gate interchangeable with their complementary counterparts,
 e.g., a locked {NAND} gate can act as {AND} or {NAND}, only depending on the key-bit.
More specifically,
one picks randomly
from the different possible configurations applicable for \TMMUX or \TMXOR, as shown in Fig.~\ref{fig:trolloc_merged}.

Continuing the example of {AND}/{NAND} gates, for \TMMUX, one would consider the 
four configurations in the upper part of Fig.~\ref{fig:trolloc_merged}(a). Note how these
configurations are pairwise indistinguishable.
Also note how the two configurations to the right are based on transforming the original gates to their counterparts---this serves for further obfuscation of the overall layout regarding the distribution
of gate types.
For \TMXOR, aside the random transformation of original gates, we also randomly select XOR/XNOR key-gates.

Concerning flip-flops (FFs), these can be locked as is, by connecting 
the Q and/or QN ports\footnote{For circuit-level efficiency, standard-cell libraries typically contain two types of FFs: those with the stored data provided as complementary pair of output signals, called Q and QN, and those
with only one output signal, either for the original data Q or the inverted data QN. Design tools then automatically instantiate, for each FF, the type which is most suitable/efficient.} to a randomly selected,
appropriate key-gate structure shown in Fig.~\ref{fig:trolloc_merged}(b, c).

\subsubsection*{No Further Information Leakage from Physical Layouts}
Aside from the fact that \TM key-gate structures are pairwise complementary, which similarly applies to most if not all locking schemes, there are \textit{no} direct correlations, neither between the types of original versus locked
	gates, nor between the locked gates and the correct key-bits.

Here it is also important to note that, for \TMMUX, all internal nets connecting to/from the MUX key-gate
share a common timing path and are, thus, jointly optimized by the synthesis engine.
Accordingly, an attacker seeking to extract any additional information from the underlying timing paths, driver
strengths, etc., cannot succeed.
For \TMXOR, this fact applies similarly as well.

\subsection{Security-and-Design-Aware Physical Synthesis}

 Next, we introduce our physical-synthesis flow that is capable of hardening any post-route
layout with \TM instances, incurring well-controlled timing and area overheads in the process.
As outlined in Fig.~\ref{fig:physical_synthesis_overall_flow}, we
start with the original netlist and other necessary data to generate a baseline layout and then carefully interleave 
locking and physical synthesis throughout two stages to produce a final protected layout.

\begin{figure}[tb]
	\centering
	\includegraphics[width=.9\columnwidth]{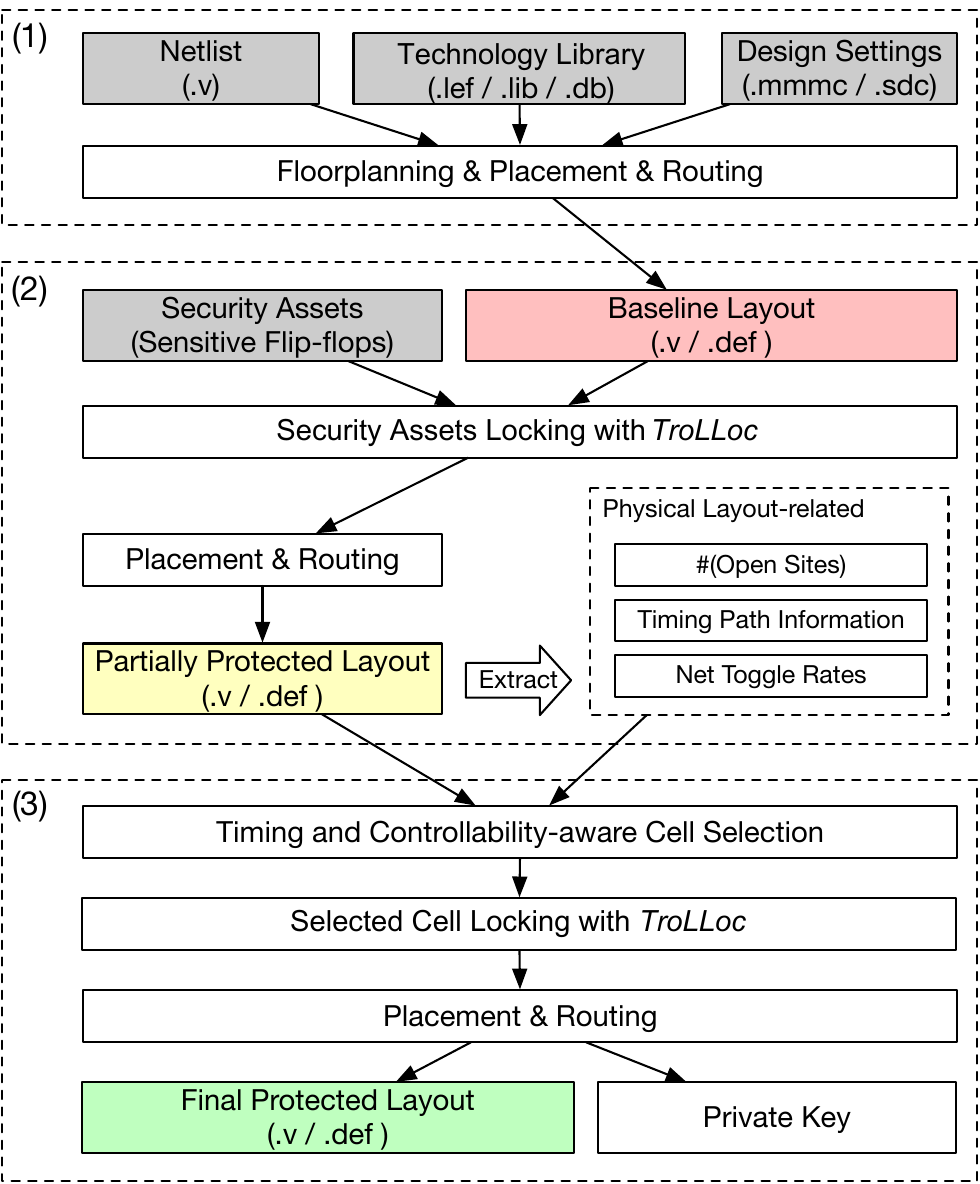}
\littlesmallerspace
	\caption{Our physical-synthesis flow.
			(1) Initial synthesis of baseline layouts. (2) Locking of security assets, followed by synthesis. (3) Locking of further components, considering
		timing and controllability, followed by synthesis.
	}
\label{fig:physical_synthesis_overall_flow}
\smallerspace
\end{figure}

\subsubsection{Two-Stage Locking Scheme}

In the first stage, we initially lock all security-critical FFs, i.e., FFs defined as assets by the user, with \TM instances.
After locking, we conduct placement and routing (P\&R), and obtain a partially protected layout (PPL).
	Note that a PPL has only assets protected; no other parts like LCNs
	are protected yet.
Further, depending on the number of user-provided assets, a PPL can still exhibit a relatively low utilization, potentially leaving the layout vulnerable to Trojan insertion.

Thus, in the second stage, we aim at locking as many more components as practically possible,
while prioritizing the
locking of LCNs, all without degrading the design quality.
Toward that end, we start by extracting some specifics obtained from the PPL: remaining open sites, timing paths, and controllability of all nets.
Then, we derive the number of cells to be locked considering open sites and timing; this step is detailed in Sec.~\ref{sec:kl}.
Next, we perform cell selection considering timing and controllability; this step is detailed in Sec.~\ref{sec:cells_sel}.
Then, all selected cells are locked. After that, P\&R is conducted again, and we retrieve a highly-utilized, final protected layout.

Finally note that initial experiments
had shown that this two-stage approach is more feasible and efficient, as in more timing-friendly, over a single-stage approach.
That is, when we did directly use the specifics extracted from the baseline layouts (marked in red in
Fig.~\ref{fig:physical_synthesis_overall_flow}) for cell selection and locking,
it was much harder to achieve timing closure, due to an accumulation of slack estimation errors. 
	As indicated, more details on the timing-and-controllability-driven cell selection in the second stage are provided in Sec.~\ref{sec:cells_sel}.

\subsubsection{Storage of Key-Bits}

For each \TM instance, there will be one corresponding key-bit, requiring some facility to store all
the key-bits.
We employ a large shift register for that, without loss of generality but for the following reasons:
\begin{itemize}
\item Prior art often considers adding a dedicated PI for each key-bit, which is not scalable.
For our, only two additional PIs are required (data in, load) for any key length.
\item Such shift register incurs negligible impact on performance and power.
For an IC locked with $k$ key-bits using \TM, loading the key will take $k$ clock cycles.
This is done once during the initial boot-up, while the main circuitry is still hold in reset.
Once the load signal is reset to \texttt{low}, the key-bits will remain stable and all related FFs
are only consuming some static power.
\item The related FFs are also helpful for
locally filling/hardening open placement sites, whereas bulky memory blocks would make this task rather complicated.
\end{itemize}

\subsubsection{On-Demand Key Length}
\label{sec:kl}
\TM instances help to occupy open placement sites with their INVs, MUXes/X(N)ORs, and FFs. 
For \TMMUX for example,
we determine the key length
for the second stage
as follows:

{\footnotesize
\begin{align}
	k = floor\left(\frac{num\_open\_sites}{size(INV) + size(MUX) + size(FF) + \alpha}\right),
	\label{eq:keylength}
\end{align}
}%
where $size(INV)$ represents the size of the smallest INV cell in the library, etc.,
and $\alpha$ is a parameter for timing budget.
For \TMXOR, we determine the key length similarly.

For $\alpha$, note that commercial IC tools conduct timing optimization by gate sizing, repeater insertion, etc. Thus, we need to reserve some space for such optimization efforts during locking. 
Accordingly, for designs where timing closure is more challenging, we will utilize a larger $\alpha$ value, and vice versa. 

\subsubsection{Timing and Controllability-Aware Cell Selection}\label{sec:cells_sel}

\begin{table}[tb]
\centering
\setlength\tabcolsep{13.0pt}
	\footnotesize
    \caption{Notations for Cell Selection}\label{tab:notations}
    \littlesmallerspace
    \begin{tabular}{c|c}
		\toprule
    {\textbf{Term}}   & {\textbf{Description}}                                 \\
		\midrule
    $C$      & The set of standard cell instances                   \\
\rowcolor{Gainsboro}
    $c$      & A cell instance from $C$                      \\
    $N$      & The set of nets                             \\ 
\rowcolor{Gainsboro}
    $n$      & A net from $N$                                \\ 
    $N(c)$   & The set of nets driven by $c$                 \\ 
\rowcolor{Gainsboro}
    $P$      & The set of timing paths                     \\ 
    $P(n)$   & The set of timing paths covering $n$          \\ 
\rowcolor{Gainsboro}
    $MS(n, P)$ & The minimum slack of paths covering $n$ \\ 
    $TPC(n)$ & The average nr.\ of toggles per clock cycle for $n$ \\
	    \bottomrule
    \end{tabular}
	\smallerspace
\end{table}

Since locking more and more components through \TM instances will
introduce further delays to all related timing paths, the selection of cells to lock becomes critical for timing closure. 
Thus, we propose a scoring function $cellScore(c, N, P)$ to comprehensively describe the priority of a cell $c$ as
follows:

{\footnotesize
\begin{align}
	cellScore(c, N, P) = &\sum_{n \in N(c)} netScore(n, P), \\
	netScore(n, P) = &\frac{1}{1 + exp(-2\cdot MS(n, P))} \cdot \frac{1}{TPC(n) + 10^{-3}}, \\
	MS(n, P) = & 
		\begin{cases}
		\min\limits_{p \in P(n)} p.slack &, \text{if } |P(n)| > 0,\\
		-0.5 &, \text{otherwise},
		\end{cases}
\end{align}
}%
with terms described in Table~\ref{tab:notations}. More details are given next.

First, the score is represented as sum of net-level scores to generalize to multi-output cases, e.g., both Q and QN of a FF are used.
Second, the net-level score is devised to
jointly consider timing and controllability. 
By using a sigmoid function, the net-level score value remains positive even for negative but small slacks.
Third, since $TPC(n) \in [0,2]$ and nets with $TPC \leq 0.1$ are considered as LCNs in this work,
we add a small margin ($10^{-3}$) to avoid both \texttt{div-by-0} and $TPC(n)$ from dominating the score.
Fourth, when calculating $MS$, some nets may not be covered by any of the reported timing paths---this scenario in general is a well-known limitation of commercial IC tools.
For those nets, $MS(n,P)$ returns -0.5 as a ``fall-back value'' which is close to observed average of negative slacks. Thus,
this value serves to conservatively penalize cells with unknown timing.

\begin{algorithm}[tb]
	\footnotesize
	\caption{
	\small
		Cell Selection Considering Timing and Controllability}\label{algo:cell_selection}
	\begin{algorithmic}[1]
		\Require Standard Cells $ C $, Nets $ N $, Timing Paths $ P $, Key Length $K$, Locking Delay $\sigma$.
		\Ensure The Set of Cells to Lock $ C' $.
		\State $C' \gets \{\}$; 
		\While{ $|C'| < K $ }
			\ForAll{cell $c \in C$} 
				\State $c.score \gets cellScore(c, N, P) $; \Comment{Score calculation for each cell}
			\EndFor
			\State Let $c_{h}$ be the cell with the highest score in $C$;
			\State $C'.append(c_{h})$;
			\State $C.remove(c_{h})$;
			
			\ForAll{net $n \in N(c_{h})$} 
				\ForAll{timing path $p \in P(n)$}
					\State $p.slack \gets p.slack - \sigma$; \Comment{Pessimistic slack estimation} 
				\EndFor
			\EndFor
		\EndWhile
		
		\State \Return $C'$;
	\end{algorithmic}
\end{algorithm}

Using the above scoring, we propose an iterative cell-selection procedure in Algorithm~\ref{algo:cell_selection}. 
There, in each iteration, we calculate the cell scores (line 3--4) and pick the cell(s) with the highest score (line 5--7). 
Next, we update the recorded slacks of all affected timing paths based on $\sigma$, i.e., a pessimistic estimate of delays introduced by \TM instances.
For \TMMUX (or \TMXOR), $\sigma$ is defined as the sum of worst-case delays for \texttt{INV\_X1} and \texttt{MUX2\_X1} (or \texttt{XOR2\_X1}), which are the default options for actual cells for \TM instances.
Note that the worst-case delays are derived for the matching corners (line 8--10).\footnote{\textit{Corners} describe different sets of parameters for characterizing an IC's operation and performance under
		varying ambient conditions.}

\section{Experimental Investigation}
\label{sec:results}
A series of thorough case studies is conducted that demonstrates the robustness, effectiveness, and efficiency of the proposed scheme.
More specifically,
in Sec.~\ref{sec:exp:ppa}, we provide a layout and security analysis, showing that \TM can serve to harden layouts
against Trojan insertion with reasonable overheads.
In Sec.~\ref{sec:exp:eco_trojan_insertion}, actual layout-level insertion of various Trojans is conducted, showing that \TM is effective and robust against such real-world attacks, which were largely overlooked in prior art.
Furthermore, in Sec.~\ref{sec:exp:2nd_MUX} and Sec.~\ref{sec:exp:2nd_XOR}, respectively,
we demonstrate the resilience of \TMMUX and \TMXOR
against second-order attacks where advanced adversaries would first try to bypass the locking defenses
before conducting Trojan insertion.

\subsection{General Setup}
\subsubsection*{Tools}

As indicated, we base our work on a commercial-grade design flow.
For experiments on \TMMUX and \TMXOR, respectively, \textit{Innovus 20.14} and \textit{Innovus 21.14} (by \textit{Cadence}) are used for physical synthesis.\footnote{The use of different versions is merely due to the fact
	that these two different sets of experiments were conducted at different points in time.
	We confirmed that only negligible differences in design quality arise, e.g., differences in utilization are within 1\%; thus, these two versions of \TM remain compatible and results as well as findings are comparable.}
The methodology is implemented in custom \textit{TCL} scripts and \textit{Python} code.

\subsubsection*{Benchmarks and Technology Library}

We employ the ISPD'22 contest benchmark suite on security closure~\cite{knechtel22_SCPL_ISPD}.
This suite contains the baseline \textit{DEF} layout files, the corresponding post-route \textit{Verilog} netlists,
the \textit{SDC} and \textit{MMMC} files used for timing analysis, the customized
\textit{LIB/LEF} files for the
well-known and publicly available
\textit{Nangate 45nm Open Cell Library}~\cite{NG45},
and custom files describing the security assets.
Since the suite was originally synthesized by some legacy versions, namely \textit{Innovus 18.13} and \textit{Innovus 16.15}, we initially resynthesize all designs at our end.
We do so for floorplan utilization rates similar to the original baseline layouts; only
the rates for \textit{AES\_3}, \textit{openMSP430\_2}, and \textit{TDEA} are set 10\% lower, as needed to lock all
security assets.

As Table~\ref{tab:bm_stat} shows, the designs vary in terms
of complexity, utilization, size (for both cells and nets), available metal layers, timing constraints, and considered corners.

\begin{table}[tb]
		\scriptsize
\setlength\tabcolsep{4.5pt}
	\centering
    \caption{Statistics for ISPD'22 Benchmarks}\label{tab:bm_stat}
\littlesmallerspace
	\begin{tabular}{c|ccccccc}
		\toprule
			\textbf{Design}        & \textbf{F. Utils} & \textbf{\#(Cells)} & \textbf{\#(Nets)} & \textbf{\#{SA}} & \textbf{\#(ML)} & \textbf{CP} & \textbf{Corner} \\
		\midrule
			AES\_1        & 75.0\%   & 16,509     & 19,694  & 291  & 10     & 1       & typical \\
\rowcolor{Gainsboro}
			AES\_2        & 75.0\%   & 16,509     & 19,694   & 291  & 10     & 1       & typical \\
			AES\_3        & 85.0\%   & 15,836     & 19,020   & 291  & 10     & 1       & typical \\
\rowcolor{Gainsboro}
			Camellia      & 50.0\%   & 6,710      & 7,160    & 256  & 6      & 10      & slow    \\
			CAST          & 50.0\%   & 12,682     & 13,057   & 192  & 6      & 10      & slow    \\
\rowcolor{Gainsboro}
			MISTY         & 50.0\%   & 9,517      & 9,904    & 204  & 6      & 10      & slow    \\
			openMSP430\_1 & 50.0\%   & 4,690      & 5,312   & 340   & 6      & 30      & slow    \\
\rowcolor{Gainsboro}
			openMSP430\_2 & 70.0\%   & 5,921      & 6,550   & 334   & 6      & 8       & slow    \\
			PRESENT       & 50.0\%   & 868       & 1,046   & 80   & 6      & 10      & slow    \\
\rowcolor{Gainsboro}
			SEED          & 50.0\%   & 12,682     & 13,057  & 195   & 6      & 10      & slow    \\
			SPARX         & 50.0\%   & 8,146      & 10,884  & 2,176   & 6      & 10      & slow    \\
\rowcolor{Gainsboro}
			TDEA          & 70.0\%   & 2,269      & 2,594   & 168   & 6      & 4       & slow  \\ 
		\bottomrule 
	\end{tabular}
	\begin{tablenotes}
		\item 
		F. Utils: floorplanning utilization for resynthesis; 
		\#(Cells/Nets): number of cells/nets in original design;
		\#(ML): number of metal layers;
		\#(SA): number of security assets;
		CP (ns): clock period;
		Corner: `typical' is characterized for 1.1V and 25C, `slow' for 0.95V and 125C.
	\end{tablenotes}
\smallerspace
\end{table}

\begin{table*}[tb]
\scriptsize
    \centering
    \caption{Layout and Security Results on the ISPD'22 Contest Benchmark Suite}\label{tab:locking_ppa_merged}
    \littlesmallerspace


    \begin{tabular}{c|c|cccccc|cccccccc}
        \toprule
        & \multirow{2}{*}{\textbf{Design}} & \multicolumn{6}{c|}{\textbf{Baseline Layout (Resynthesized)}} & \multicolumn{8}{c}{\textbf{Protected Layout (Final)}}                                                                   \\[0.5mm]
        & & \textit{Utils}       & \textit{\#(Open)}  & \textit{TU}   & \textit{WNS}        & \textit{TNS}        & \textit{Power}     & \textit{Utils}   & \textit{\#(Open)} & \textit{$\Delta$(Open)} & \textit{TU} &
		\textit{WNS}    & \textit{TNS}    & \textit{Power}  & \textit{KL} \\
        \midrule
	& AES\_1        & 75.4\% & 43,980 & 8.7\% & 0.000  & 0.000  & 59.957 & 96.2\% & 6,838 & -84.5\% & 11.1\% & -0.013 & -0.043 & 60.552 & 1,199 \\
\rowcolor{Gainsboro} \cellcolor{white}
        & AES\_2        & 75.1\% & 44,420 & 8.7\% & -0.001 & -0.003 & 59.441 & 96.3\% & 6,649 & -85.0\% & 10.9\% & -0.008 & -0.017 & 61.040 & 1,202 \\
        & AES\_3        & 85.7\% & 22,129 & 9.7\% & -0.001 & -0.002 & 59.869 & 96.6\% & 5,225 & -76.4\% & 11.2\% & -0.031 & -4.657 & 62.787 & 441   \\
\rowcolor{Gainsboro} \cellcolor{white}
        & Camellia      & 49.5\% & 33,919 & 9.5\% & 1.194  & 0.000  & 1.233  & 98.9\% & 753   & -97.8\% & 13.3\% & 0.015  & 0.000  & 1.488  & 1,271 \\
        & CAST          & 49.1\% & 54,444 & 9.1\% & 0.047  & 0.000  & 3.136  & 93.6\% & 6,879 & -87.4\% & 12.7\% & -0.134 & -1.455 & 3.912  & 1,572 \\
\rowcolor{Gainsboro} \cellcolor{white}
        & MISTY         & 48.5\% & 43,359 & 8.6\% & -0.021 & -0.037 & 2.238  & 94.4\% & 4,686 & -89.2\% & 12.4\% & -0.140 & -1.515 & 2.966  & 1,215 \\
        & openMSP430\_1 & 49.7\% & 32,799 & 6.2\% & 0.000  & 0.000  & 0.375  & 97.9\% & 1,389 & -95.8\% & 12.6\% & 0.000  & 0.000  & 0.544  & 1,218 \\
\rowcolor{Gainsboro} \cellcolor{white}
        & openMSP430\_2 & 70.5\% & 15,125 & 7.7\% & 0.000  & 0.000  & 1.239  & 97.1\% & 1,497 & -90.1\% & 10.6\% & -0.006 & -0.015 & 1.369  & 543   \\
        & PRESENT       & 49.8\% & 6,284  & 4.4\% & 6.694  & 0.000  & 0.198  & 98.0\% & 245   & -96.1\% & 6.8\%  & 4.935  & 0.000  & 0.235  & 241   \\
\rowcolor{Gainsboro} \cellcolor{white}
        & SEED          & 49.1\% & 54,444 & 9.1\% & 0.047  & 0.000  & 3.136  & 94.3\% & 6,056 & -88.9\% & 12.7\% & -0.182 & -1.072 & 3.908  & 1,612 \\
        & SPARX         & 49.9\% & 69,979 & 6.3\% & 2.452  & 0.000  & 2.164  & 98.8\% & 1,658 & -97.6\% & 14.0\% & 0.027  & 0.000  & 2.822  & 2,582 \\
\rowcolor{Gainsboro} \cellcolor{white}
        \multirow{-12}{*}{\rotatebox[origin=c]{90}{\textbf{\TMMUX}}} 
        & TDEA          & 70.4\% & 5,559  & 6.8\% & 0.049  & 0.000  & 1.084  & 98.4\% & 304   & -94.5\% & 8.0\%  & 0.027  & 0.000  & 1.153  & 214  \\
        \midrule    
	& AES\_1                  & 74.9\% & 44,781   & 8.5\% & 0.000  & -0.001 & 58.671 & 96.1\% & 6,922    & -84.5\%     & 10.7\% & -0.014 & -0.160 & 59.689 & 1,200 \\
\rowcolor{Gainsboro} \cellcolor{white}
        & AES\_2                  & 74.7\% & 45,152   & 8.5\% & 0.000  & 0.000  & 58.766 & 95.0\% & 8,971    & -80.1\%     & 10.6\% & -0.007 & -0.057 & 59.554 & 1,149 \\
        & AES\_3                  & 86.3\% & 21,244   & 9.8\% & -0.004 & -0.011 & 59.727 & 95.5\% & 6,921    & -67.4\%     & 10.7\% & -0.003 & -0.042 & 60.719 & 416   \\
\rowcolor{Gainsboro} \cellcolor{white}
        & Camellia                & 49.5\% & 33,930   & 9.2\% & 0.856  & 0.000  & 1.237  & 98.7\% & 864      & -97.5\%     & 12.2\% & 0.075  & 0.000  & 1.478  & 1,348 \\
        & CAST                    & 49.1\% & 54,501   & 8.9\% & 0.023  & 0.000  & 3.143  & 94.6\% & 5,774    & -89.4\%     & 11.7\% & -0.094 & -0.934 & 3.867  & 1,804 \\
\rowcolor{Gainsboro} \cellcolor{white}
        & MISTY                   & 48.5\% & 43,365   & 8.7\% & -0.013 & -0.013 & 2.266  & 93.9\% & 5,102    & -88.2\%     & 11.8\% & -0.097 & -1.554 & 2.945  & 1,315 \\
        & openMSP430\_1           & 49.6\% & 32,813   & 6.1\% & 0.000  & 0.000  & 0.375  & 98.7\% & 875      & -97.3\%     & 10.1\% & 0.000  & 0.000  & 0.542  & 1,337 \\
\rowcolor{Gainsboro} \cellcolor{white}
        & openMSP430\_2           & 70.4\% & 15,170   & 7.7\% & -0.004 & -0.008 & 1.238  & 97.4\% & 1,349    & -91.1\%     & 10.0\% & 0.000  & 0.000  & 1.371  & 582   \\
        & PRESENT                 & 49.8\% & 6,274    & 4.2\% & 5.654  & 0.000  & 0.197  & 97.4\% & 327      & -94.8\%     & 6.2\%  & 4.994  & 0.000  & 0.234  & 255   \\
\rowcolor{Gainsboro} \cellcolor{white}
        & SEED                    & 49.1\% & 54,501   & 8.9\% & 0.023  & 0.000  & 3.143  & 94.1\% & 6,364    & -88.3\%     & 11.6\% & -0.024 & -0.106 & 3.745  & 1,802 \\
        & SPARX                   & 49.8\% & 70,038   & 6.2\% & 2.389  & 0.000  & 2.161  & 99.0\% & 1,419    & -98.0\%     & 13.3\% & 0.319  & 0.000  & 2.888  & 2,746 \\
\rowcolor{Gainsboro} \cellcolor{white}
        \multirow{-12}{*}{\rotatebox[origin=c]{90}{\textbf{\TMXOR}}}
        & TDEA                    & 70.0\% & 5,634    & 6.8\% & 0.021  & 0.000  & 1.083  & 99.0\% & 185      & -96.7\%     & 8.1\%  & 0.017  & 0.000  & 1.173  & 226   \\
        \bottomrule    

    \end{tabular}

    \begin{tablenotes}
        \item
	Utils: utilization after physical synthesis;
	\#(Open): nr.\  of open sites;
        TU: track utilization;
        WNS (ns): worst negative slack;
		TNS (ns): total negative slack;
        Power (mW): total power;
        $\Delta$(Open): reduction of nr.\  of open sites;
        KL: key length (bits), accounting both for locked assets and other locked cell instances.

	\end{tablenotes}
    \smallerspace
\end{table*}

\begin{figure*}[tb]
	\captionsetup[subfigure]{labelformat=empty}
		\centering
		\subfloat[]{\includegraphics[width=.32\linewidth, height=4cm]{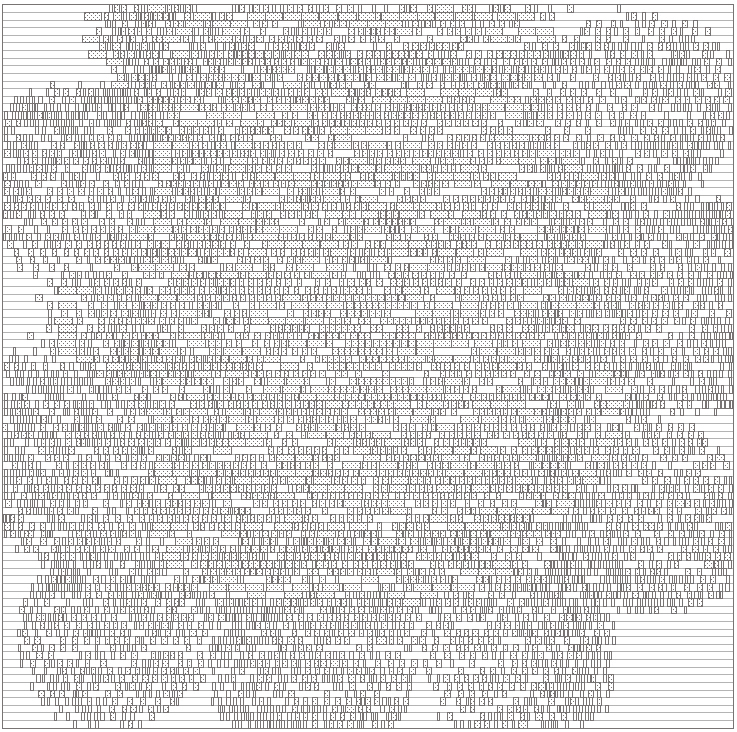}
		\label{fig:pre_logic_locking_demo} 		
		}
		\hfill
		\subfloat[]{\includegraphics[width=.32\linewidth, height=4cm]{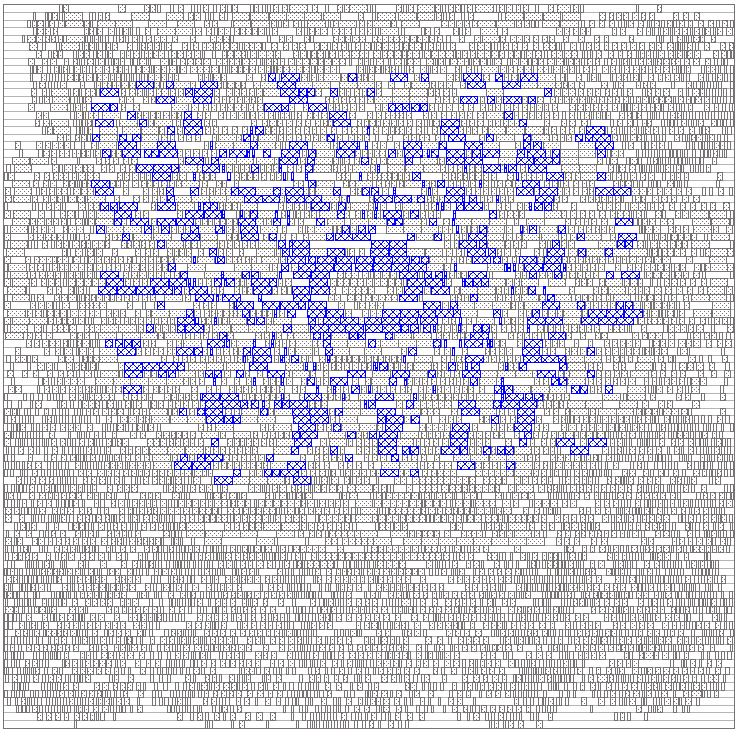}
	\label{fig:post_lock_asset_demo}
		}
		\hfill
		\subfloat[]{\includegraphics[width=.32\linewidth, height=4cm]{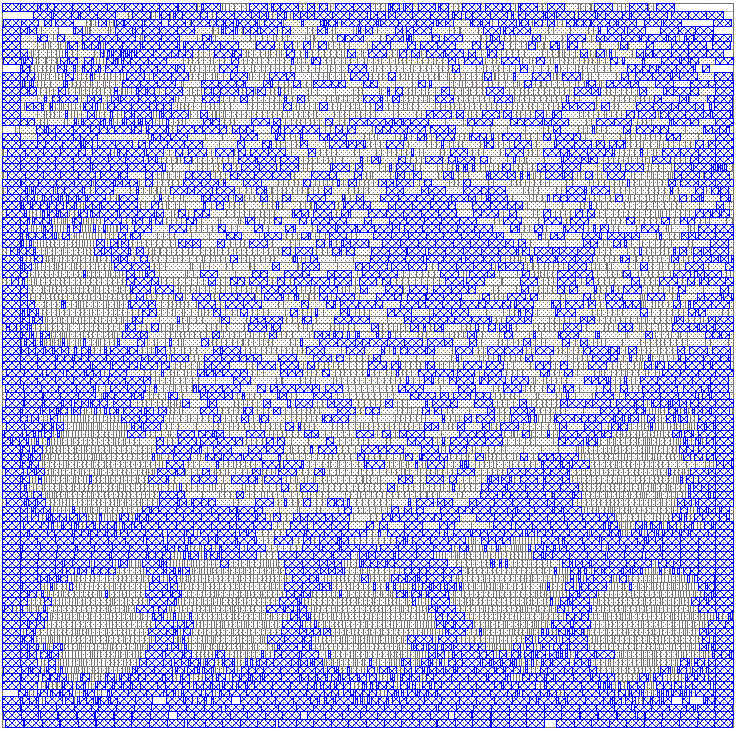}
	\label{fig:post_lock_lcn_demo}
		}
\littlesmallerspace
\littlesmallerspace
\littlesmallerspace
		\caption{Layout illustrations for the benchmark \textit{Camellia} under \TMMUX:
			(left) baseline layout, i.e., after initial resynthesis;
			(middle) after first-stage locking of security assets,
			(right) final protected layout, i.e., after second-stage locking of other cells considering timing and controllability.
			Cells introduced by \TMMUX instances are marked in blue, whereas others are marked in grey. 
			The utilization increases from 49.5\% (left) to 59.2\% (middle), and eventually to 98.9\% (right);
				also recall Footnote~\ref{fn:util}.
			Note that layouts heights are reduced here; the original aspect ratio is 1.0.
		}
		\label{fig:logic_locking_demo}
\smallerspace
\end{figure*}

\subsection{Case Study 1: ISPD'22 Benchmarks}\label{sec:exp:ppa}

We quantify security and layout results for the
resynthesized baseline layouts versus the final, protected layouts in Table~\ref{tab:locking_ppa_merged}.
In Fig.~\ref{fig:logic_locking_demo}, we provide an example for \TMMUX applied on the \textit{Camellia} benchmark.
More details are discussed next.

First, all the protected layouts exhibit much higher utilization rates: 93.6--99.0\%.\footnote{%
	\label{fn:util}%
	In general, the achievable utilization is subject to the IC design and
	the technology library.
	For our work, showcasing such ultra-high utilization is important---it supports our claim of truly hardening layouts against Trojan insertion. As indicted, prior art like~\cite{xiao14} is challenged by high utilization.}
Recall that, unlike prior art, we do not utilize any fillers or spares toward that end.
This achievement helps to \ul{significantly reduce open placement sites}, namely by 90.3\% and 89.5\% on average, respectively, for \TMMUX and \TMXOR.
In turn, \ul{this renders all layouts more resilient against Trojan insertion}---the less
open sites remain, the more difficult Trojan insertion becomes~\cite{knechtel22_SCPL_ISPD,trippel20}.
We also achieve higher routing track utilization (defined as \textit{total routed wire length} / \textit{total track length});
thus, \ul{routing for Trojans will become more challenging as well}~\cite{knechtel22_SCPL_ISPD,trippel20}.

Second, it is essential to note that, \ul{despite the much higher utilization rates achieved without using any filler or spares, all protected layouts are without DRC violations}.
This proves the efficiency and real-world relevance of our proposed scheme, in particular of the EDA flow for security closure.
In fact, we can tune the flow to achieve this very trade-off: largest possible utilization without introducing DRC violations.
Doing so is a well-known challenge for IC design in general~\cite{baek22}, and our method is fully supporting the security-concerned IC designer with automated guidance and feedback.

When considering the two findings above together, the following important \textbf{research question} arises: \textit{facing such layout hardening, can adversaries still succeed in exploiting the remaining open sites for Trojan insertion, or will they be hindered by DRC
violations and/or other real-world design challenges?} To answer this, as indicated, we conduct actual Trojan insertion on the hardened layouts
in Sec.~\ref{sec:exp:eco_trojan_insertion}.

Finally, for overheads induced by \TM, we note the following.
Total power is increased by 18.5\% for \TMMUX and by 17.7\% for \TMXOR on average, respectively. This is reasonable given all the additional cells introduced through \TM instances.
Concerning performance, the average and the maximal worst negative slacks (WNS)\footnote{%
WNS quantifies the worst timing path,
i.e.,
the path with the largest negative value and, thus, the path that exhibits the largest timing violation. Any WNS number is to be interpreted in context of the clock period.}
are 1.41\% and 3.10\% for \TMMUX, and 0.54\% and 1.40\% for \TMXOR, respectively. This is reasonable as well, again, given all the additional \TM circuitry.
It is also important to note that, unlike DRCs or other critical design checks, such WNS violations do \textit{not} render the IC unfit, neither for manufacturing nor for actual use. The only requirement for such ICs
would be to operate them with an accordingly reduced clock period.

\subsection{Case Study 2: ISPD'23 Trojan Insertion}\label{sec:exp:eco_trojan_insertion}

\subsubsection*{Setup}

Here we also employ parts of the ISPD'23 suite~\cite{eslami23}, namely the actual Trojans.
	Note that we refrain from using the ISPD'23 suite in full here---for a fair evaluation of the attackers prospects,
	i.e., to put the related outcomes for Trojan insertion correctly into the context of the prior security and
	layout analysis in Sec.~\ref{sec:exp:ppa}, we have to continue working with the ISPD'22 benchmarks here.

Only a subset of Trojans are readily compatible with the ISPD'22 suite, namely those targeting on the very same cell instances for the respective benchmarks.
We obtain all 9 related Trojans by performing technology mapping at our end.
Also note that we employ the very same ECO techniques for Trojan insertion as in the ISPD'23 contest; we thank the ISPD'23 organizers for sharing the related scripts with us.

\subsubsection*{Results}

\begin{table*}[tb]
\scriptsize
\setlength\tabcolsep{3.8pt}
    \centering
    \caption{Results for ISPD'23 Trojan Insertion on the ISPD'22 Contest Benchmark Suite}\label{tab:ISPD23}
    \littlesmallerspace


    \begin{tabular}{c|c|ccccccc|ccccccc}
        \toprule
        & & \multicolumn{7}{c|}{\textbf{Baseline Layout (Resynthesized)}} & \multicolumn{7}{c}{\textbf{Protected Layout (Final)}}
	\\
	& {\textbf{Design -- Trojan / \{Trojans\}}}
		& \multicolumn{2}{c}{\textbf{Mode `reg'}}
		& \multicolumn{2}{c}{\textbf{Mode `adv'}}
		& \multicolumn{2}{c}{\textbf{Mode `adv2'}}
		& {\textbf{\textit{Res}}}
		& \multicolumn{2}{c}{\textbf{Mode `reg'}}
		& \multicolumn{2}{c}{\textbf{Mode `adv'}}
		& \multicolumn{2}{c}{\textbf{Mode `adv2'}}
		& {\textbf{\textit{Res}}}
	\\[0.5mm]
        &
	& \textit{\#(Vios)}       & \textit{TNS}        
	& \textit{\#(Vios)}       & \textit{TNS}        
	& \textit{\#(Vios)}       & \textit{TNS}        
	&
	& \textit{\#(Vios)}       & \textit{TNS}        
	& \textit{\#(Vios)}       & \textit{TNS}        
	& \textit{\#(Vios)}       & \textit{TNS}
	&
	\\
        \midrule

	& Camellia -- \{burn random, burn targeted\}	& 0  & 0.000  & 0 & 0.000  & 0 & 0.000  & \sym{no} & 11 & 0.000  & 21 & 0.000  & 89    & -3.022  & \symwide{_no_} \\
\rowcolor{Gainsboro} \cellcolor{white}
	& Camellia -- fault random	& 28 & 0.000  & 0 & 0.000  & 0 & 0.000  & \sym{no} & 95 & 0.000  & 33 & 0.000  & 26    & -0.189  & \symwide{_yes_} \\
	& Camellia -- fault targeted	& 19 & 0.000  & 0 & 0.000  & 0 & 0.000  & \sym{no} & 54 & -0.102 & 28 & 0.000  & 51    & -2.611  & \symwide{_yes_} \\
\rowcolor{Gainsboro} \cellcolor{white}
	& Camellia -- \{leak random, leak targeted\}	& 0  & 0.000  & 0 & 0.000  & 0 & 0.000  & \sym{no} & 62 & 0.000  & 61 & 0.000  & 2,737 & -99.210 & \symwide{_yes_} \\
	& SEED -- burn random		& 0  & -3.265 & 0 & -3.293 & 0 & -3.184 & \symwide{_yes_} & 0  & -2.249 & 0  & -2.224 & 40    & -2.395  & \symwide{_yes_}~\sym{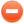} \\
\rowcolor{Gainsboro} \cellcolor{white}
	& SEED -- fault random		& 12 & -3.633 & 0 & -3.624 & 0 & -3.353 & \symwide{_yes_} & 35 & -2.168 & 18 & -2.326 & 24    & -2.045  & \symwide{_yes_}~\sym{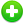} \\
\multirow{-7}{*}{\rotatebox[origin=c]{90}{\textbf{\TMMUX}}} 
	& SEED -- leak random		& 0  & -3.177 & 0 & -3.188 & 0 & -3.290 & \symwide{_yes_} & 0  & -2.137 & 0  & -2.199 & 11    & -4.370  & \symwide{_yes_}~\sym{-} \\

        \midrule    

\rowcolor{Gainsboro} \cellcolor{white}
	& Camellia -- \{burn random, burn targeted\}	& 0  & 0.000  & 0 & 0.000  & 0 & 0.000  & \sym{no} & \multicolumn{2}{c}{\sym{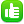}} & \multicolumn{2}{c}{\sym{pos}} & 146 & -0.222 & \symwide{_yes_} \\
	& Camellia -- fault random	& 40 & 0.000  & 0 & 0.000  & 0 & 0.000  & \sym{no} & \multicolumn{2}{c}{\sym{pos}} & \multicolumn{2}{c}{\sym{pos}} & 164 & -0.769 & \symwide{_yes_} \\
\rowcolor{Gainsboro} \cellcolor{white}
	& Camellia -- fault targeted	& 20 & 0.000  & 0 & 0.000  & 0 & 0.000  & \sym{no} & \multicolumn{2}{c}{\sym{pos}} & \multicolumn{2}{c}{\sym{pos}} & 95  & -0.002 & \symwide{_yes_} \\
	& Camellia -- \{leak random, leak targeted\}	& 0  & 0.000  & 0 & 0.000  & 0 & 0.000  & \sym{no} & \multicolumn{2}{c}{\sym{pos}} & \multicolumn{2}{c}{\sym{pos}} & 300 & -2.531 & \symwide{_yes_} \\
\rowcolor{Gainsboro} \cellcolor{white}
	& SEED -- burn random		& 0  & -1.188 & 0 & -1.312 & 0 & -1.170 & \symwide{_no_} & 0   & -0.750 & 3   & -0.798 & 10  & -7.123 & \symwide{_no_}~\sym{-} \\
	& SEED -- fault random		& 14 & -1.683 & 0 & -1.543 & 0 & -0.938 & \symwide{_no_} & 25  & -0.728 & 6   & -0.887 & 10  & -1.790 & \symwide{_no_}~\sym{+} \\
\rowcolor{Gainsboro} \cellcolor{white}
\multirow{-7}{*}{\rotatebox[origin=c]{90}{\textbf{\TMXOR}}} 
	& SEED -- leak random		& 0  & -1.301 & 0 & -1.163 & 0 & -1.430 & \symwide{_no_} & 16  & -0.770 & 13  & -0.802 & 10  & -3.100 & \symwide{_no_}~\sym{+} \\

        \bottomrule    

    \end{tabular}

    \begin{tablenotes}
        \item Mode: Trojan insertion mode as in ISPD'23 contest;
	\#(Vios):~nr.\ of DRC and/or other design-check violations; 
        TNS (ns):~total negative slack;
	\sym{pos}:~Trojan insertion failed entirely, specifically due to exceedingly high utilization after insertion.
	Res:~interpretation of final result, i.e., best-case outcome for adversaries across all three modes, as follows --
	\sym{no}:~Trojan insertion succeeded without any violations;
	\symwide{_no_}:~Trojan insertion succeeded with managable violations / \TM failed partially;
	\symwide{_yes_}:~Trojan insertion succeded but only with excessive violations / \TM mainly successful.
	Note that \TM interpretations are only applicable for protected layouts.
	Besides, note that, for cases where the outcomes are identical across different Trojan instances, only one row is listed, with the corresponding Trojans being provided in \{\} set notation.
	Finally note that, for cases where the final results' interpretation is the same for the baseline and protected layouts, we further annotate whether violations have
		\sym{+} increased or \sym{-} decreased for Trojan insertion into the protected layout over insertion into the baseline layout.
	\end{tablenotes}
    \smallerspace
\end{table*}

We quantify the violation outcomes after Trojan insertion in Table~\ref{tab:ISPD23}, thereby judging the effectiveness of \TM versus the practicality of threats arising from layout-level Trojan insertion.\footnote{%
	\label{fn:TI_results_interpret}
Note the following for interpretation of Table~\ref{tab:ISPD23}. Considering the number of DRC and/or other design-check violations, and the total negative slacks (TNS), we derive a conservative interpretation of the
	final result, i.e., the best-case outcome with least violations and smallest TNS value across all three different modes for Trojan insertion as in the ISPD'23 contest. The threshold between `manageable' versus
	`excessive' violations are, following some rules of thumb from the real-world of IC design, set to $>25$ DRC violations and TNS values exceeding 20\% of the clock period. For cases where both the baseline and the protected layouts lead to the same final
	result, we further compare the degree of violations, to judge whether \TM managed to incur more or less challenges for the adversary toward final design closure after Trojan insertion. Toward that end, DRC
	violations are prioritized over TNS violations. This is because most if not all DRC violations must be fixed by adversary in order to release the design for actual manufacturing without raising suspicions or
	even failures, whereas \textit{some} TNS violations may be kept, as these would `merely' force the IC to only operate correctly for a reduced clock period---such outcome may well be blamed on process variations.
	Note that we focus here on TNS instead of WNS; TNS serves better to quantify the efforts imposed onto the adversaries for possibly fixing all timing issues, whereas WNS only quantifies by how much the whole IC
	violates timing.
}
In Fig.~\ref{fig:HT_example}, we provide some examples.

Overall, for the 9 Trojans in total, actual insertion was made more difficult in 7 case for \TMMUX and in 8 cases for \TMXOR, respectively, and made even impractical in 4 and 6 out of those cases where excessive violations are induced
on the attackers' efforts.
More details are discussed next.

\begin{figure}[tb]
	\captionsetup[subfigure]{labelformat=empty}
		\centering
		\includegraphics[width=.9\columnwidth, height=4.8cm, draft=false]{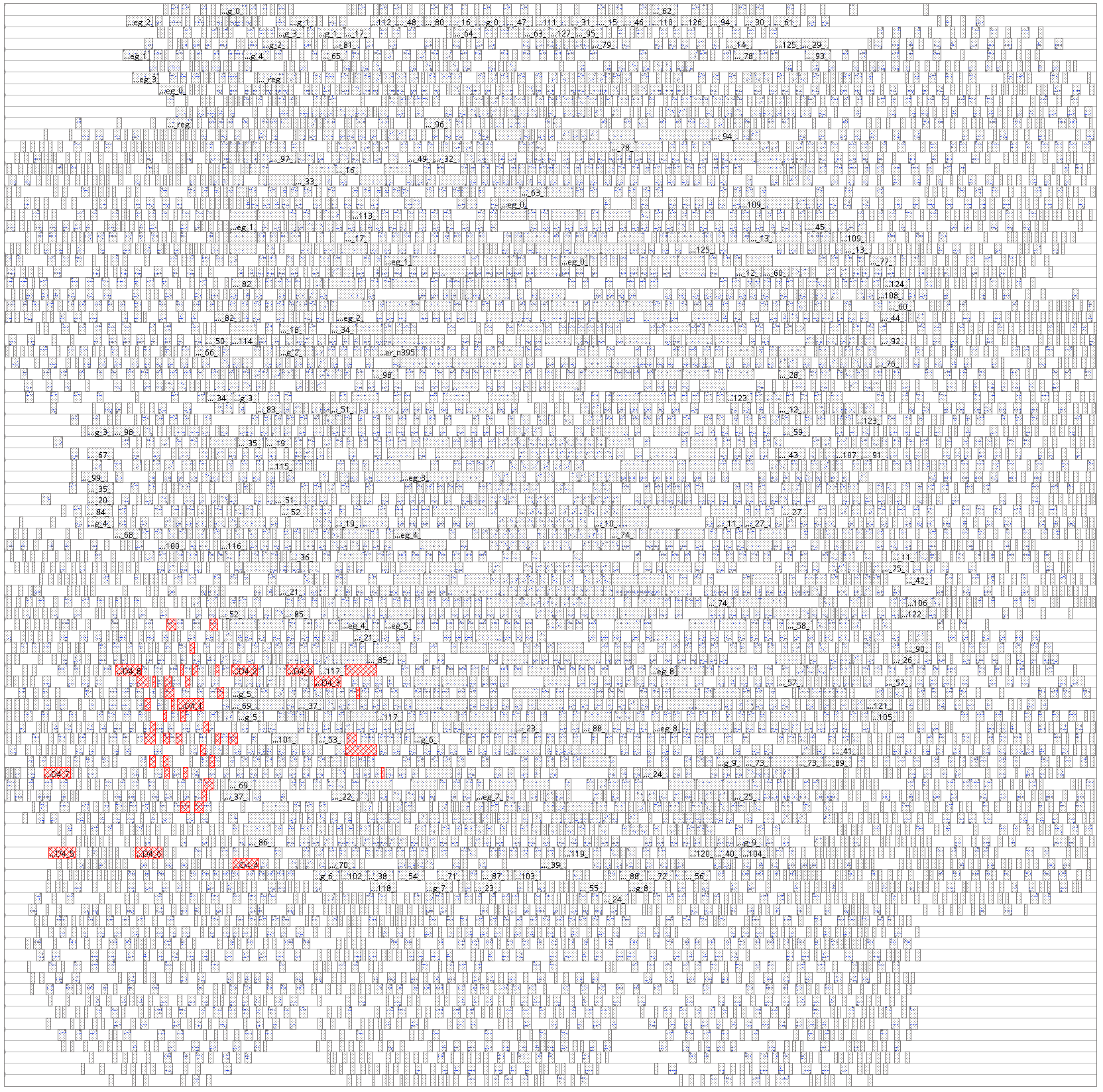}\\[0.75em]
		~\includegraphics[width=.9\columnwidth, height=4.8cm, draft=false]{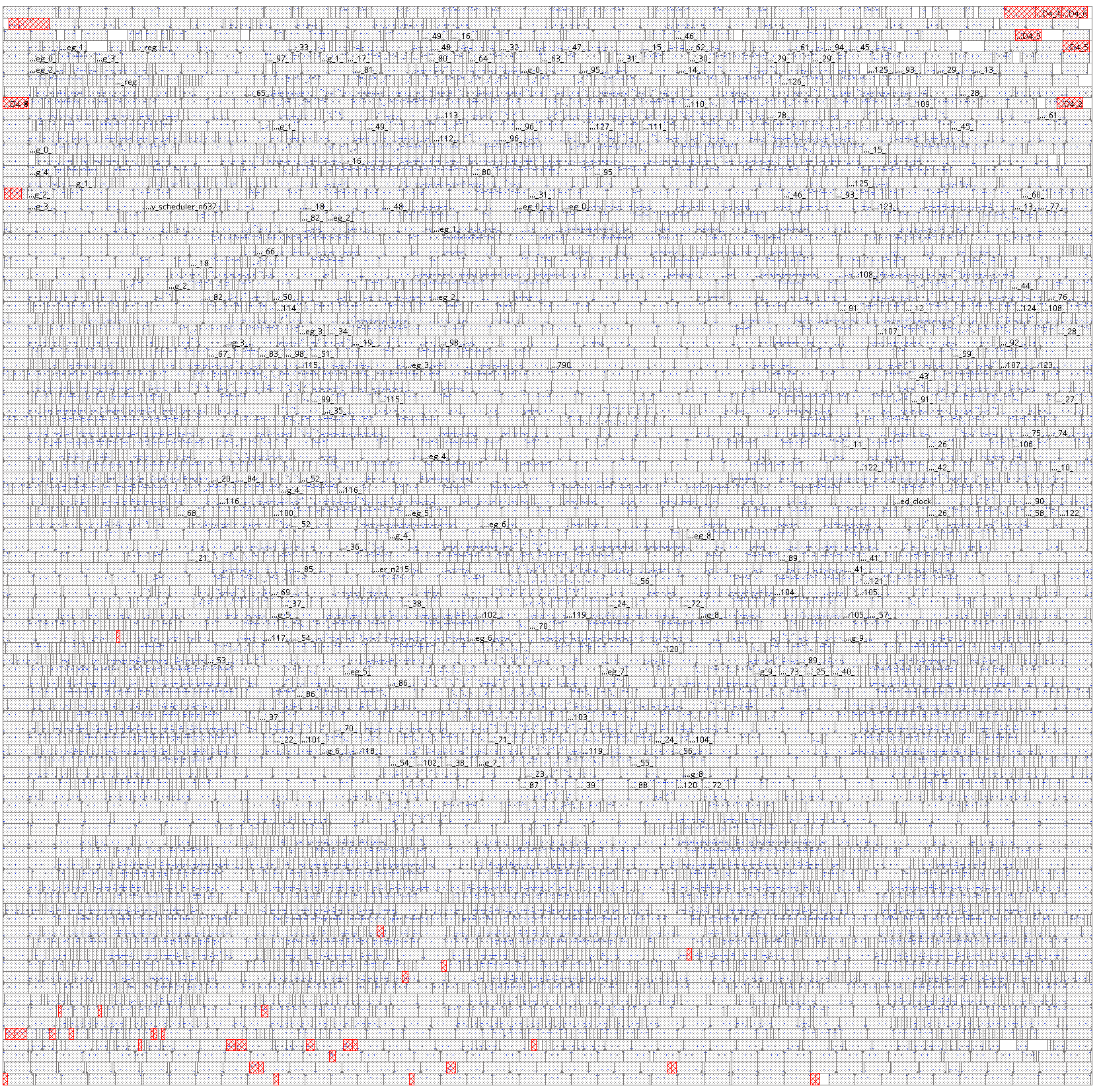}
\littlesmallerspace
		\caption{Layout illustrations for the benchmark \textit{Camellia}, after the Trojan \textit{leak targeted} is inserted using the ISPD'23 `reg' mode.
			(Top) Trojan inserted into the baseline layout.
			(Bottom) Trojan inserted into the final layout as protected by \TMMUX.
			All benign cells, including \TMMUX instances, are marked in grey, whereas Trojan cells are marked in red.
			Note that layouts heights are reduced here; the original aspect ratio is 1.0.
		}
		\label{fig:HT_example}
\smallerspace
\end{figure}

First, \ul{\textit{TroLLoc} in general is particularly effective} for \textit{Camellia}, i.e., a benchmark \ul{where ultra-high utilization is achieved for the protected layouts}.
Some related layout examples are shown in Fig.~\ref{fig:HT_example}: while Trojan insertion was successful without any violations for the baseline layout, it effectively failed due to excessive
violations for the protected layout. For the latter, also compare to Fig.~\ref{fig:logic_locking_demo}, to see how the Trojan logic only barely fitted into the few remaining open sites. Due to the resulting spread-out nature
of the Trojan placement, excessive timing violations did arise; due to the ultra-high utilization, a large number of DRC violations did arise as well.

Second, \ul{\textit{TroLLoc-XOR} is slightly more effective than \textit{TroLLoc-MUX}.}
Interestingly, the layouts protected by \TMXOR exhibit little lower utilization numbers (Table~\ref{tab:locking_ppa_merged}).
This implies that \ul{layout hardening against post-design insertion of actual Trojans is more intricate than `only' requiring to push for ultra-high utilization}. This finding argues the converse of earlier prior art
like~\cite{ba16,knechtel21_SCPL_ICCAD,xiao14}, but aligns with more recent and more practical works like~\cite{wei24}.

Third, although \TM cannot prevent Trojan insertion altogether, it is still effective.
Here it is important to note that, while not directly comparable due to different technology setups, \ul{none of the winning teams of the ISPD'23 contest could fully prevent Trojan insertion
either}~\cite{eslami23}.
This implies that \ul{post-design insertion of actual Trojans, specifically through commercial-grade ECO techniques, represents a severe threat}. This finding aligns with recent prior art~\cite{perez22,AlexTRJ,wei24}.

Fourth, while not explicitly quantified here,
it is important to note that, even in cases where Trojan insertion
succeeded with manageable violations, the adversaries would still face two challenges: (i)~to resolve the violations, which can become more challenging than usual for these ultra-dense layouts, and (ii)~to be
able to utilize the Trojan later on in the field.
For (i), note that adversaries would have to tackle most if not all DRC violations and also keep TNS violations within tight bounds; also recall Footnote~\ref{fn:TI_results_interpret}.
For (ii), note that each and every security asset targeted at by 
Trojans has to be accounted for its underlying obfuscation, i.e., all the assets' signal values could be either as is or inverted, depending on the unknown locking key.
Therefore, although not explicitly quantified here either, all the actual Trojan implementations would likely become more complex and larger and, thus, more difficult to insert.

To summarize and answer the research question raised in Sec.~\ref{sec:exp:ppa}: 
\ul{adversaries are considerably but not entirely hindered by \textit{TroLLoc} for post-design Trojan insertion, whereas an assessment of actual prospects
requires case-by-case studies}. From the perspective of commercial-grade IC design, this is expected: any SOTA IC design requires thorough efforts even for regular, security-agonistic research and development.

\subsection{Case Study 3: Second-Order Attacks on \TMMUX}\label{sec:exp:2nd_MUX}

Here we evaluate \TMMUX against SOTA ML-based attacks that specifically target on MUX-based locking: \ML~\cite{alrahis22_ML} and \SC~\cite{alaql21}.
We also utilize a SOTA structural attack, \RS~\cite{almeida23}.

\subsubsection*{Setup for \ML}

We obtain the \ML implementation from~\cite{alrahis22_ML} and configure it as follows.
First, we extend the one-hot feature vectors to capture all possible gates utilized in the ISPD'22 suite.
Second, consistent with the original operation,
{we represent each gate, PI, and PO as a separate node in a graph, with all connections between gates and ports represented as edges.}
Note that the connections between the input/output of \TMMUX key-gates are kept as test set for prediction, while all others
are used as training set.
Finally, we adopt the same graph neural-network configuration and training
hyperparameters as in~\cite{alrahis22_ML}; specifically, the number of hops while extracting subgraphs is set to 3 and
the threshold for post-processing is set to 0.01.

\subsubsection*{Setup for \SC and \RSC}

We obtain the \SC implementation from~\cite{alaql21} and the \RSC implementation from~\cite{RSC}.
Note that
\RS employs majority voting for many \SC runs conducted over a large range of resynthesized netlists,
which are structurally different yet functionally equivalent to the design under attack. Thus, we refer to this setup as \RSC.

We configure both tools as follows.
First, since \SC cannot handle any loops in the designs, e.g., introduced through sequential feedback operations on FFs,
we convert any FFs into pseudo PIs/POs.
Note that this is the technically correct handling of sequential designs as unrolled combinational designs, and doing so is common practice in the literature.
Also note that, for the FFs building up the shift register storing the key, we designate all related pseudo PIs as individual key-bit PIs.
Second, we utilize the \RSC framework with its default settings; the framework then automatically handles all required steps, like conversion from \textit{Verilog} to \textit{bench} netlists as required for
\SC, etc.

\subsubsection*{Metrics}
We are using the following established metrics: \textit{accuracy (AC)}, \textit{precision (PC)}, and \textit{key prediction accuracy (KPA)}.
AC measures the percentage of correctly deciphered key-bits, i.e., $(k_{correct}/k_{total})$.
PC measures the correctly deciphered key-bits, optimistically considering every $X$/undeciphered value as a correct guess, i.e., $((k_{correct} + k_{X})/k_{total})$.
KPA measures the correctly deciphered key-bits over the entire prediction set, i.e., $(k_{correct}/(k_{total}-k_{X}))$.
Furthermore, we consider the \textit{constant propagation effect (COPE)}, which specifically measures the vulnerability against the \SC attack;
in short, COPE = 0\% means the attack fails entirely.
All metrics are reported in percentage.

\subsubsection{Results for \ML}

\begin{table}[tb]
\scriptsize
\setlength\tabcolsep{3.8pt}
    \centering
    \caption{Results for \ML on \TMMUX and \textit{D-MUX}}
    \label{tab:muxlink_result}
\littlesmallerspace
        \begin{tabular}{c|cccc|cccc}
	\toprule
            \multirow{2}{*}{\textbf{Design}} & \multicolumn{4}{c|}{\textbf{\TMMUX}}                               & \multicolumn{4}{c}{\textbf{\textit{D-MUX}}}            \\ 
                                    & \textit{AC} & \textit{PC} & \textit{KPA} & \textit{\#(X)} & \textit{AC} & \textit{PC} & \textit{KPA} & \textit{\#(X)} \\ 
	\midrule
            AES\_1                  & 28.7\%   & 71.4\%   & 50.2\%    & 512   & 78.1\%   & 79.9\%   & 79.5\%    & 21    \\
\rowcolor{Gainsboro} 
            AES\_2                  & 28.1\%   & 75.0\%   & 52.9\%    & 564   & 80.4\%   & 81.7\%   & 81.4\%    & 15    \\
            AES\_3                  & 08.1\%   & 90.0\%   & 45.0\%    & 361   & 88.4\%   & 89.1\%   & 89.0\%    & 3     \\
\rowcolor{Gainsboro} 
            Camellia                & 26.1\%   & 73.8\%   & 49.9\%    & 606   & 90.6\%   & 92.6\%   & 92.4\%    & 25    \\
            CAST                    & 37.4\%   & 66.0\%   & 52.4\%    & 449   & 93.7\%   & 94.3\%   & 94.2\%    & 7     \\
\rowcolor{Gainsboro} 
            MISTY                   & 33.3\%   & 65.8\%   & 49.3\%    & 395   & 97.8\%   & 98.0\%   & 98.0\%    & 3     \\
            openMSP430\_1           & 23.5\%   & 76.6\%   & 50.1\%    & 646   & NA  & NA   & NA   & NA    \\
\rowcolor{Gainsboro} 
            openMSP430\_2           & 09.3\%   & 90.4\%   & 49.5\%    & 440   & 66.8\%   & 69.9\%   & 69.0\%    & 17    \\
            PRESENT                 & 11.6\%   & 95.0\%   & 70.0\%    & 201   & NA  & NA   & NA   & NA    \\
\rowcolor{Gainsboro} 
            SEED                    & 37.0\%   & 64.0\%   & 50.7\%    & 435   & 97.3\%   & 97.7\%   & 97.7\%    & 6     \\
            SPARX                   & 05.3\%   & 94.9\%   & 51.4\%    & 2,312 & NA   & NA   & NA   & NA    \\
\rowcolor{Gainsboro} 
            TDEA                    & 01.8\%   & 99.0\%   & 66.6\%    & 208   & NA   & NA   & NA   & NA    \\ 
	\bottomrule
            \end{tabular}

    \begin{tablenotes}
        \item 
AC: accuracy;
PC: precision;
KPA: key prediction accuracy;
\#(X): nr.\ of undeciphered key-bits;
NA: locking using the script in~\cite{alrahis22_ML} fails.
\end{tablenotes}
    \smallerspace
\end{table}

The results in Table~\ref{tab:muxlink_result} show that \ul{\textit{MuxLink} can predict only 20.85\% of all key-bits correctly}, on average.
Interestingly, it holds across the various benchmarks (except for \textit{SPARX}) that, the larger is the key length, the higher is \ML's accuracy.
However, this does \textit{no} imply that designs hardened with larger \TMMUX keys would be more prone to second-order attacks. This is not the case because, at the same time,
(1) \ul{the average KPA is 53.17\% which shows that, overall, \textit{MuxLink} is not performing much better than random guessing}, and (2) that average KPA remains rather steady across designs.
In other words, for designs with larger \TMMUX keys, while \ML succeeds to correctly infer a larger share of key-bits, it also infers an equally larger share of key-bits incorrectly.
Thus, \ul{the scalability of \textit{TroLLoc-MUX}'s robustness against \textit{MuxLink} is confirmed.}

For the sake of putting \ML's performance on \TMMUX into some context for prior art,
we have implemented \textit{D-MUX}~\cite{sisejkovic22,alrahis22_ML} and configured it for for the same key length as \TMMUX (Table~\ref{tab:locking_ppa_merged}).
The corresponding results,
also shown in Table~\ref{tab:muxlink_result}, clearly demonstrate that \ul{\textit{TroLLoc-MUX} is superior to \textit{D-MUX}, across all metrics.}

\subsubsection{Results for \SC}

\begin{table*}[tb]
\scriptsize
\setlength\tabcolsep{5.2pt}
    \centering
    \caption{Results for \SC and \RSC on \TMMUX}
    \label{tab:resynth_scope}
\littlesmallerspace
        \begin{tabular}{c|ccccccc|ccccccc}
	\toprule
            \multirow{2}{*}{\textbf{Design}} & \multicolumn{7}{c|}{\textbf{\textit{SCOPE}}}                               & \multicolumn{7}{c}{\textbf{\textit{Resynthesis + SCOPE}}}            \\ 
                                    & \textit{AC(1)} & \textit{PC(1)} & \textit{KPA(1)} & \textit{AC(2)} & \textit{PC(2)} & \textit{KPA(2)} & \textit{COPE}
                                    & \textit{AC(1)} & \textit{PC(1)} & \textit{KPA(1)} & \textit{AC(2)} & \textit{PC(2)} & \textit{KPA(2)} & \textit{Avg. COPE}
				    \\ 
	\midrule

AES\_1		&	21.10\%	& 79.23\%	& 50.39\%	&	20.76\%	& 78.89\%	& 49.60\%	& 18.45\%	&	32.77\%	& 68.14\%	& 50.70\%	&	31.85\%	& 67.22\%	& 49.29\%	&  17.89\%	\\
\rowcolor{Gainsboro}
AES\_2		&	22.46\%	& 78.03\%	& 50.56\%	&	21.96\%	& 77.53\%	& 49.43\%	& 17.87\%	&	33.94\%	& 65.97\%	& 49.93\%	&	34.02\%	& 66.05\%	& 50.06\%	&  17.13\%	\\
AES\_3		&	9.52\%	& 90.92\%	& 51.21\%	&	9.07\%	& 90.47\%	& 48.78\%	& 10.32\%	&	12.24\%	& 86.16\%	& 46.95\%	&	13.83\%	& 87.75\%	& 53.04\%	&  9.96\%	\\
\rowcolor{Gainsboro}
Camellia	&	11.40\%	& 91.42\%	& 57.08\%	&	8.57\%	& 88.59\%	& 42.91\%	& 0.10\%	&	34.46\%	& 68.13\%	& 51.95\%	&	31.86\%	& 65.53\%	& 48.04\%	&  0.23\%	\\
CAST		&	12.40\%	& 87.91\%	& 50.64\%	&	12.08\%	& 87.59\%	& 49.35\%	& 0.05\%	&	41.92\%	& 58.96\%	& 50.53\%	&	41.03\%	& 58.07\%	& 49.46\%	&  0.17\%	\\
\rowcolor{Gainsboro}
MISTY		&	17.77\%	& 81.15\%	& 48.53\%	&	18.84\%	& 82.22\%	& 51.46\%	& 0.35\%	&	40.49\%	& 57.94\%	& 49.05\%	&	42.05\%	& 59.50\%	& 50.94\%	&  0.53\%	\\
openMSP430\_1	&	16.09\%	& 84.07\%	& 50.25\%	&	15.92\%	& 83.90\%	& 49.74\%	& 0.24\%	&	27.99\%	& 72.41\%	& 50.36\%	&	27.58\%	& 72.00\%	& 49.63\%	&  0.65\%	\\
\rowcolor{Gainsboro}
openMSP430\_2	&	13.25\%	& 88.76\%	& 54.13\%	&	11.23\%	& 86.74\%	& 45.86\%	& 0.18\%	&	23.02\%	& 81.03\%	& 54.82\%	&	18.96\%	& 76.97\%	& 45.17\%	&  0.95\%	\\
PRESENT		&	1.24\%	& 94.60\%	& 18.75\%	&	5.39\%	& 98.75\%	& 81.25\%	& 2.79\%	&	18.25\%	& 75.51\%	& 42.71\%	&	24.48\%	& 81.74\%	& 57.28\%	&  2.40\%	\\
\rowcolor{Gainsboro}
SEED		&	11.97\%	& 87.22\%	& 48.37\%	&	12.77\%	& 88.02\%	& 51.62\%	& 0.05\%	&	40.63\%	& 58.62\%	& 49.54\%	&	41.37\%	& 59.36\%	& 50.45\%	&  0.16\%	\\
SPARX		&	4.02\%	& 96.51\%	& 53.60\%	&	3.48\%	& 95.97\%	& 46.39\%	& 0.14\%	&	7.39\%	& 92.40\%	& 49.35\%	&	7.59\%	& 92.60\%	& 50.64\%	&  0.16\%	\\
\rowcolor{Gainsboro}
TDEA		&	0.46\%	& 100\%		& 100\%		&	0.00\%	& 99.53\%	& 0.00\%	& 0.01\%	&	10.28\%	& 86.91\%	& 44.00\%	&	13.08\%	& 89.71\%	& 56.00\%	&  0.60\%	\\

	\bottomrule
            \end{tabular}

    \begin{tablenotes}
        \item 
AC: accuracy;
PC: precision;
KPA: key prediction accuracy;
COPE: constant propagation effect~\cite{alaql21}.
\end{tablenotes}
    \smallerspace
\end{table*}

The results in Table~\ref{tab:resynth_scope} show that
\ul{\textit{SCOPE} can predict only 11.81\% / 11.67\%\footnote{%
\SC predicts two mutually complementary keys. The related metrics are reported as `AC(1)' versus `AC(2)' etc., respectively.}
of all key-bits correctly}, on average.
\ul{The average KPA of 52.79\% / 47.20\% shows that \textit{SCOPE} is largely equivalent to random guessing} as well.
Note that we observe similar trends for key length and accuracy with \SC as with \ML; 
\ul{the scalability of \textit{TroLLoc-MUX}'s robustness against \textit{SCOPE} is confirmed.}
Also note that \SC can correctly decipher only a single bit for the benchmark {TDEA}, which trivially results in $PC(1) = KPA(1) = 100\%$ and $PC(2) \approx 100\%, KPA(2) = 0\%$.

\subsubsection{Results for \RSC}

The results in Table~\ref{tab:resynth_scope} show that \ul{\textit{Resynthesis} \textit{+} \textit{SCOPE} can predict 26.95\% / 27.31\% of all key-bits correctly}, on average.
While these accuracy results are superior to those of \SC alone---as expected---this does \textit{not} mean that \TMMUX is more prone to second-order attacks by \RSC.
This is because \ul{the average KPA of 49.16\% / 50.83\% shows that \textit{Resynthesis} \textit{+} \textit{SCOPE} does not perform any better than random guessing either.}
Similar trends for key length and accuracy are observed again;
\ul{the scalability of \textit{TroLLoc-MUX}'s robustness against \textit{Resynthesis} \textit{+} \textit{SCOPE} is confirmed as well.}

\subsection{Case Study 4: Second-Order Attacks on \TMXOR}\label{sec:exp:2nd_XOR}

\subsubsection*{Setup}

Here we evaluate \TMXOR against \OM~\cite{alrahis21_OMLA},
a SOTA attack that specifically targets on XOR-based locking.
We obtain the \OM implementation from~\cite{alrahis21_OMLA}.
Note that, unlike \SC or \textit{Muxlink}, \OM always provides an inference for all key-bits. Thus,
$k_{X} = 0$ and $AC = PC = KPA$.

We configure this case study as follows.
First, note that, in its original implementation, when given a locked design to attack, \OM seeks to lock further parts of that design and then utilizes those newly locked parts as training data.
For \TMXOR, however, any further locking is impractical, as the protected layouts are already locked as much as possible, exhibiting ultra-high utilization and/or very few relevant nets remain unlocked.
Thus, for this case study, we train \OM via a classical leave-one-out scheme, considering all respectively remaining, locked designs from the ISPD'22 suite.

Second, based on the findings for \TMXOR without gate balancing (Sec.~\ref{sec:TMXOR_wo_bal}),
we revise the cell selection described in Algorithm~\ref{algo:cell_selection} as follows.
For each gate type, we track the number of locked instances. During the iterative selection and locking procedure, we then balance each type with its complementary counterpart, e.g., \texttt{NAND2} and \texttt{AND2}.
Before each round of cell selection, we temporarily ignore a cell if locking that cell would break the balance.

\subsubsection{Results for \TMXOR without Gate Balancing}
\label{sec:TMXOR_wo_bal}

\begin{table}[tb]
\scriptsize
\setlength\tabcolsep{7.0pt}
    \centering
    \caption{Results for \OM on \TMXOR}
    \littlesmallerspace
    \label{tab:omla_result}
        \begin{tabular}{c|ccc|ccc}
	\toprule
            \multirow{2}{*}{\textbf{Design}} & \multicolumn{3}{c|}{\textbf{\TMXOR, Unbalanced}}  & \multicolumn{3}{c}{\textbf{\TMXOR, Balanced}} \\ 
                                    & \textit{KL}    & \textit{Utils} & \textit{KPA} & \textit{KL}         & \textit{Utils} & \textit{KPA} \\
	\midrule
            AES\_1                  & 1,200 & 96.1\%       & 79.1\%    & 1,273      & 95.0\%            & 51.5\%         \\
\rowcolor{Gainsboro} 
            AES\_2                  & 1,149 & 95.0\%       & 78.9\%    & 1,215      & 94.3\%            & 52.1\%         \\
            AES\_3                  & 416   & 95.5\%       & 63.7\%    & 416        & 95.6\%            & 52.6\%         \\
\rowcolor{Gainsboro} 
            Camellia                & 1,348 & 98.7\%       & 75.6\%    & 1,051      & 86.8\%            & 48.8\%         \\
            CAST                    & 1,804 & 94.6\%       & 78.2\%    & 1,707      & 96.2\%            & 68.4\%         \\
\rowcolor{Gainsboro} 
            MISTY                   & 1,315 & 93.9\%       & 78.9\%    & 1,172      & 93.8\%            & 67.6\%         \\
            openMSP430\_1           & 1,337 & 98.7\%       & 70.2\%    & 1,154      & 91.2\%            & 56.8\%         \\
\rowcolor{Gainsboro} 
            openMSP430\_2           & 582   & 97.4\%       & 66.5\%    & 582        & 97.2\%            & 55.0\%         \\
            PRESENT                 & 255   & 97.4\%       & 47.8\%    & 208        & 88.7\%            & 52.9\%         \\
\rowcolor{Gainsboro} 
            SEED                    & 1,802 & 94.1\%       & 78.5\%    & 1,618      & 92.4\%            & 66.9\%         \\
            SPARX                   & 2,746 & 99.0\%       & 56.6\%    & 2,741      & 98.9\%            & 48.2\%         \\
\rowcolor{Gainsboro} 
            TDEA                    & 226   & 99.0\%       & 54.0\%    & 226        & 99.0\%            & 53.5\%         \\
	\bottomrule
            \end{tabular}
	\begin{tablenotes}
		\item 
		KL: key length (bits), accounting both for locked assets and other locked cell instances;
		Utils: utilization after physical synthesis;
		KPA: key prediction accurracy.
	\end{tablenotes}
    \smallerspace
    \end{table}

The results in Table~\ref{tab:omla_result} show that, \ul{without gate balancing being applied for \textit{TroLLoc-XOR}, \textit{OMLA} can achieve an average KPA of 69\%, which is notably better than random guessing.}
This is due to the fact that the composition of locked gates can be unbalanced in terms of the original, unlocked gate's type. 
For example, in \textit{AES\_1}, 441 \texttt{NAND2} gates are locked while there are only 37 \texttt{AND2} gates locked for \TMXOR without balancing applied.
Such trends occur in other designs as well, from which \OM learns about the majority of the original gates' types.
That is, for the same example of \textit{AES\_1}, whenever \OM seeks to decipher a \TMXOR key-gate that is locked via \texttt{NAND2} or \texttt{AND2} gates, it will predict the original gate to be a \texttt{NAND2} with much higher probability.

\subsubsection{Results for \TMXOR with Gate Balancing}

The results in Table~\ref{tab:omla_result} show that, \ul{with gate balancing being applied for \textit{TroLLoc-XOR}, \textit{OMLA} can only achieve an average KPA of 56.19\%, which
is much closer to random guessing} than \textit{TroLLoc-XOR} without gate balancing, namely by 12.81 percentage points (pps).
For the previous example of \textit{AES\_1}, KPA is even reduced by 27.60 pps.

We note that \TMXOR with gate balancing applied can be limited---by the original gate composition---in terms of achievable utilization and key length.
In fact, key lengths are 5.76\% shorter and utilization numbers are 2.53 pps lower on average.
To mitigate this limitation, guided efforts are required from the IC-design stage prior to physical synthesis, i.e., logic synthesis;\footnote{%
Gate balancing could become another optimization objective, in addition to the classical
power, performance, and area (PPA) objectives.
Similar efforts for security-aware logic synthesis already exist,
e.g., to reduce power side-channel information leakage by means of synthesizing differential logic~\cite{knechtel20_Sec_EDA_DATE}.
}
such efforts are left for future work.

\section{Conclusion}
\label{sec:conc}

We propose a novel scheme for proactive, pre-silicon Trojan prevention by means of logic locking and layout hardening, all in a real-world setup for IC design.
Our work successfully addresses three key challenges for related prior art and for Trojan countermeasures in general:
	(i)~ours remains robust against potential second-order attacks, i.e., adversaries seeking to bypass the layout-level defense before actual Trojan insertion,
	(ii)~ours is effective as in protecting layouts in general and security assets in particular against real-world insertion of various Trojans,
	and (iii)~ours is efficient as in achieving favorable security-versus-cost trade-offs, by careful integration of our techniques into commercial-grade IC tools.

\bibliographystyle{IEEEtran}
\bibliography{main}

\begin{thebibliography}{10}
\providecommand{\url}[1]{#1}
\csname url@samestyle\endcsname
\providecommand{\newblock}{\relax}
\providecommand{\bibinfo}[2]{#2}
\providecommand{\BIBentrySTDinterwordspacing}{\spaceskip=0pt\relax}
\providecommand{\BIBentryALTinterwordstretchfactor}{4}
\providecommand{\BIBentryALTinterwordspacing}{\spaceskip=\fontdimen2\font plus
\BIBentryALTinterwordstretchfactor\fontdimen3\font minus
  \fontdimen4\font\relax}
\providecommand{\BIBforeignlanguage}[2]{{%
\expandafter\ifx\csname l@#1\endcsname\relax
\typeout{** WARNING: IEEEtran.bst: No hyphenation pattern has been}%
\typeout{** loaded for the language `#1'. Using the pattern for}%
\typeout{** the default language instead.}%
\else
\language=\csname l@#1\endcsname
\fi
#2}}
\providecommand{\BIBdecl}{\relax}
\BIBdecl

\bibitem{Wang23}
F.~Wang \emph{et~al.}, ``Security closure of {IC} layouts against hardware
  {Trojans},'' in \emph{Proc. ISPD}, 2023.

\bibitem{release}
------, ``Baseline and protected layouts,'' 2024,
  \url{https://drive.google.com/file/d/17ysAYnoPgFuy0Yj969dTGUWZVrIT7PmW/view?usp=sharing}.

\bibitem{rostami14}
M.~Rostami \emph{et~al.}, ``A primer on hardware security: Models, methods, and
  metrics,'' \emph{Proc. IEEE}, vol. 102, no.~8, 2014.

\bibitem{wu21}
W.~{Hu} \emph{et~al.}, ``An overview of hardware security and trust: Threats,
  countermeasures and design tools,'' \emph{TCAD}, 2020.

\bibitem{knechtel21_Sec_Emerg_ISPD}
J.~Knechtel, ``Hardware security for and beyond {CMOS} technology,'' in
  \emph{Proc. ISPD}, 2021.

\bibitem{knechtel20_Sec_EDA_DATE}
J.~Knechtel \emph{et~al.}, ``Towards secure composition of integrated circuits
  and electronic systems: On the role of {EDA},'' in \emph{Proc. DATE}, 2020.

\bibitem{sigl11}
G.~Sigl, ``Keynote address: Design of secure systems -- where are the eda
  tools?'' in \emph{Proc. ICCAD}, 2011.

\bibitem{ravi19}
P.~Ravi \emph{et~al.}, ``Security is an architectural design constraint,''
  \emph{Microprocess. Microsyst.}, vol.~68, no.~C, 2019.

\bibitem{knechtel22_SCPL_ISPD}
J.~Knechtel \emph{et~al.}, ``Benchmarking security closure of physical layouts:
  Ispd 2022 contest,'' in \emph{Proc. ISPD}, 2022.

\bibitem{knechtel21_SCPL_ICCAD}
------, ``Security closure of physical layouts,'' in \emph{Proc. ICCAD}, 2021.

\bibitem{eslami23}
M.~Eslami \emph{et~al.}, ``Benchmarking advanced security closure of physical
  layouts: Ispd 2023 contest,'' 2023, \url{https://wp.nyu.edu/ispd23_contest/}.

\bibitem{xiao16}
K.~Xiao \emph{et~al.}, ``Hardware trojans: Lessons learned after one decade of
  research,'' \emph{TODAES}, vol.~22, no.~1, 2016.

\bibitem{dong20}
C.~Dong \emph{et~al.}, ``Hardware trojans in chips: A survey for detection and
  prevention,'' \emph{Sensors}, vol.~20, no.~18, 2020.

\bibitem{yang16_a2}
K.~Yang \emph{et~al.}, ``{A2}: Analog malicious hardware,'' in \emph{Proc. SP},
  2016.

\bibitem{YRS20}
M.~Yasin \emph{et~al.}, \emph{Trustworthy Hardware Design: Combinational Logic
  Locking Techniques}.\hskip 1em plus 0.5em minus 0.4em\relax Springer, 2020.

\bibitem{alrahis22_ML}
L.~Alrahis \emph{et~al.}, ``Muxlink: Circumventing learning-resilient
  mux-locking using graph neural network-based link prediction,'' in
  \emph{Proc. DATE}, 2022.

\bibitem{alrahis21_OMLA}
------, ``Omla: An oracle-less machine learning-based attack on logic
  locking,'' \emph{TCS}, vol.~69, no.~3, 2022.

\bibitem{almeida23}
F.~Almeida \emph{et~al.}, ``Resynthesis-based attacks against logic locking,''
  in \emph{Proc. ISQED}, 2023.

\bibitem{alaql21}
A.~Alaql \emph{et~al.}, ``Scope: Synthesis-based constant propagation attack on
  logic locking,'' \emph{Trans. VLSI}, vol.~29, no.~8, 2021.

\bibitem{perez22}
T.~Perez and S.~Pagliarini, ``Hardware {Trojan} insertion in finalized layouts:
  From methodology to a silicon demonstration,'' \emph{TCAD}, 2022.

\bibitem{adee08}
S.~Adee, ``The hunt for the kill switch,'' \emph{Spectrum}, vol.~45, no.~5,
  2008.

\bibitem{Skorobogatov12}
S.~Skorobogatov and C.~Woods, ``Breakthrough silicon scanning discovers
  backdoor in military chip,'' in \emph{Proc. TCHES}, 2012.

\bibitem{muehlberghuber13}
M.~Muehlberghuber \emph{et~al.}, ``Red team vs. blue team hardware {Trojan}
  analysis: Detection of a hardware {Trojan} on an actual {ASIC},'' in
  \emph{Proc. Int. Workshop Hardw. Arch. Supp. Sec. Priv.}, 2013.

\bibitem{ghandali20}
S.~Ghandali \emph{et~al.}, ``Side-channel hardware {Trojan} for provably-secure
  {SCA}-protected implementations,'' \emph{Trans. VLSI}, vol.~28, no.~6, pp.
  1435--1448, 2020.

\bibitem{Puschner23}
E.~Puschner \emph{et~al.}, ``Red team vs. blue team: A real-world hardware
  {Trojan} detection case study across four modern {CMOS} technology
  generations,'' in \emph{Proc. SP}, 2023.

\bibitem{trippel20}
T.~Trippel \emph{et~al.}, ``Icas: an extensible framework for estimating the
  susceptibility of ic layouts to additive trojans,'' in \emph{Proc. SP}, 2020.

\bibitem{salmani16}
H.~Salmani \emph{et~al.}, ``Vulnerability analysis of a circuit layout to
  hardware trojan insertion,'' \emph{Trans. Inf. Forens. Sec.}, vol.~11, no.~6,
  2016.

\bibitem{AlexTRJ}
A.~Hepp \emph{et~al.}, ``A pragmatic methodology for blind hardware {Trojan}
  insertion in finalized layouts,'' in \emph{Proc. ICCAD}, 2022.

\bibitem{wei24}
X.~Wei \emph{et~al.}, ``Rethinking ic layout vulnerability: Simulation-based
  hardware trojan threat assessment with high fidelity,'' in \emph{Proc. SP},
  2024.

\bibitem{dupuis14}
S.~Dupuis \emph{et~al.}, ``A novel hardware logic encryption technique for
  thwarting illegal overproduction and hardware trojans,'' in \emph{Proc.
  IOLTS}.

\bibitem{marcelli17}
A.~Marcelli \emph{et~al.}, ``An evolutionary approach to hardware encryption
  and trojan-horse mitigation,'' in \emph{Proc. DATE}, 2017.

\bibitem{samimi16}
M.~S. Samimi \emph{et~al.}, ``Hardware enlightening: No where to hide your
  hardware trojans!'' in \emph{Proc. IOLTS}, 2016.

\bibitem{sisejkovic19}
D.~\v{S}i\v{s}ejkovi\'{c} \emph{et~al.}, ``Control-lock: Securing processor
  cores against software-controlled hardware trojans,'' in \emph{Proc.
  GLSVLSI}, 2019.

\bibitem{xiao14}
K.~Xiao \emph{et~al.}, ``A novel built-in self-authentication technique to
  prevent inserting hardware trojans,'' \emph{TCAD}, vol.~33, no.~12, 2014.

\bibitem{ba15}
P.-S. Ba \emph{et~al.}, ``Hardware trojan prevention using layout-level design
  approach,'' in \emph{Proc. Europ, Conf. Circ. Theory Des.}, 2015.

\bibitem{ba16}
------, ``Hardware trust through layout filling: A hardware trojan prevention
  technique,'' in \emph{Proc. ISVLSI}, 2016.

\bibitem{hosseintalaee17}
H.~Hossein-Talaee and A.~Jahanian, ``Layout vulnerability reduction against
  trojan insertion using security-aware white space distribution,'' in
  \emph{Proc. ISVLSI}, 2017.

\bibitem{imeson13}
F.~Imeson \emph{et~al.}, ``Securing computer hardware using {3D} integrated
  circuit ({IC}) technology and split manufacturing for obfuscation,'' in
  \emph{Proc. USENIX Sec. Symp.}, 2013.

\bibitem{shi18}
Q.~Shi \emph{et~al.}, ``Obfuscated built-in self-authentication with secure and
  efficient wire-lifting,'' \emph{TCAD}, vol.~38, no.~11, 2018.

\bibitem{trippel23}
T.~Trippel \emph{et~al.}, ``{T-TER}: Defeating {A2} {Trojans} with targeted
  tamper-evident routing,'' in \emph{Proc. CCS}, 2023.

\bibitem{Guo23}
G.~Guo \emph{et~al.}, ``{ASSURER}: A {PPA}-friendly security closure framework
  for physical design,'' in \emph{Proc. ASP-DAC}, 2023.

\bibitem{Hsu23}
J.-W. Hsu \emph{et~al.}, ``Security-aware physical design against {Trojan}
  insertion, frontside probing, and fault injection attacks,'' in \emph{Proc.
  ISPD}, 2023.

\bibitem{Eslami23a}
M.~Eslami \emph{et~al.}, ``{SALSy}: Security-aware layout synthesis,'' 2023.

\bibitem{wei23}
X.~Wei \emph{et~al.}, ``Gdsii-guard: Eco anti-trojan optimization with
  exploratory timing-security trade-offs,'' in \emph{Proc. DAC}, 2023.

\bibitem{Eslami24}
M.~Eslami \emph{et~al.}, ``{SCARF}: Securing chips with a robust framework
  against fabrication-time hardware trojans,'' 2024.

\bibitem{guimaraes17}
L.~A. Guimar{\~a}es \emph{et~al.}, ``Detection of layout-level trojans by
  monitoring substrate with preexisting built-in sensors,'' in \emph{Proc.
  ISVLSI}, 2017.

\bibitem{hou18}
Y.~Hou \emph{et~al.}, ``R2d2: Runtime reassurance and detection of a2 trojan,''
  in \emph{Proc. HOST}, 2018.

\bibitem{vijayan20}
A.~Vijayan \emph{et~al.}, ``Runtime identification of hardware trojans by
  feature analysis on gate-level unstructured data and anomaly detection,''
  \emph{TODAES}, vol.~25, no.~4, 2020.

\bibitem{kim11_trojan}
L.~W. Kim and J.~D. Villasenor, ``A system-on-chip bus architecture for
  thwarting integrated circuit trojan horses,'' \emph{Trans. VLSI}, 2011.

\bibitem{basak17}
A.~Basak \emph{et~al.}, ``Security assurance for system-on-chip designs with
  untrusted {IPs},'' \emph{Trans. Inf. Forens. Sec.}, vol.~12, no.~7, 2017.

\bibitem{wu16}
T.~F. Wu \emph{et~al.}, ``{TPAD}: Hardware trojan prevention and detection for
  trusted integrated circuits,'' \emph{TCAD}, vol.~35, no.~4, 2016.

\bibitem{wahby16}
R.~S. Wahby \emph{et~al.}, ``Verifiable {ASICs},'' \emph{Proc. SP}, 2016.

\bibitem{Agrawal07}
D.~Agrawal \emph{et~al.}, ``{Trojan} detection using {IC} fingerprinting,'' in
  \emph{Proc. SP}, 2007.

\bibitem{guo19_QIFV}
X.~{Guo} \emph{et~al.}, ``{QIF-Verilog}: Quantitative information-flow based
  hardware description languages for pre-silicon security assessment,'' in
  \emph{Proc. HOST}, 2019.

\bibitem{fern17}
N.~Fern \emph{et~al.}, ``Detecting hardware trojans in unspecified
  functionality through solving satisfiability problems,'' in \emph{Proc.
  ASP-DAC}, 2017.

\bibitem{chen18}
X.~Chen \emph{et~al.}, ``Hardware trojan detection in third-party digital
  intellectual property cores by multilevel feature analysis,'' \emph{TCAD},
  2018.

\bibitem{Lashen23}
H.~Lashen \emph{et~al.}, ``{TrojanSAINT}: Gate-level netlist sampling-based
  inductive learning for hardware {Trojan} detection,'' in \emph{Proc. ISCAS},
  2023.

\bibitem{narasimhan13}
S.~Narasimhan \emph{et~al.}, ``Hardware trojan detection by multiple-parameter
  side-channel analysis,'' \emph{Trans. Comp.}, vol.~62, no.~11, 2013.

\bibitem{deng20}
D.~Deng \emph{et~al.}, ``Novel design strategy toward a2 trojan detection based
  on built-in acceleration structure,'' \emph{TCAD}, vol.~39, no.~12, 2020.

\bibitem{sugawara14}
T.~Sugawara \emph{et~al.}, ``Reversing stealthy dopant-level circuits,''
  \emph{J. Cryptogr. Eng.}, vol.~5, no.~2, 2015.

\bibitem{vashistha18}
N.~Vashistha \emph{et~al.}, ``Trojan scanner: Detecting hardware trojans with
  rapid {SEM} imaging combined with image processing and machine learning,'' in
  \emph{ISTFA}, 2018.

\bibitem{gohil22_ATT}
V.~Gohil \emph{et~al.}, ``{ATTRITION}: Attacking static hardware {Trojan}
  detection techniques using reinforcement learning,'' in \emph{Proc. CCS},
  2022.

\bibitem{kento17}
K.~Hasegawa \emph{et~al.}, ``{Trojan}-feature extraction at gate-level netlists
  and its application to hardware-{Trojan} detection using random forest
  classifier,'' in \emph{Proc. ISCAS}, 2017.

\bibitem{jacob14}
N.~Jacob \emph{et~al.}, ``Hardware {Trojans}: Current challenges and
  approaches,'' \emph{Comp. \& Dig. Techs.}, vol.~8, no.~6, 2014.

\bibitem{baek22}
K.~Baek \emph{et~al.}, ``Pin accessibility and routing congestion aware {DRC}
  hotspot prediction using graph neural network and u-net,'' in \emph{Proc.
  ICCAD}, 2022.

\bibitem{roy10}
J.~A. Roy \emph{et~al.}, ``Ending piracy of integrated circuits,''
  \emph{Computer}, vol.~43, no.~10, 2010.

\bibitem{8429401}
H.~Mardani~Kamali \emph{et~al.}, ``Lut-lock: A novel lut-based logic
  obfuscation for fpga-bitstream and asic-hardware protection,'' in \emph{Proc.
  ISVLSI}, 2018.

\bibitem{rajendran15}
J.~Rajendran \emph{et~al.}, ``Fault analysis-based logic encryption,''
  \emph{Trans. Comp.}, vol.~64, no.~2, 2015.

\bibitem{yasin16_test}
M.~Yasin \emph{et~al.}, ``Activation of logic encrypted chips: Pre-test or
  post-test?'' in \emph{Proc. DATE}, 2016.

\bibitem{Subramanyan2015}
P.~Subramanyan \emph{et~al.}, ``Evaluating the security of logic encryption
  algorithms,'' in \emph{Proc. HOST}, 2015.

\bibitem{chakraborty18}
P.~Chakraborty \emph{et~al.}, ``Sail: Machine learning guided structural
  analysis attack on hardware obfuscation,'' in \emph{Proc. AHOST}, 2018.

\bibitem{NG45}
``{NanGate FreePDK45 Open Cell Library},'' Nangate Inc, 2011,
  \url{http://www.nangate.com/?page_id=2325}.

\bibitem{RSC}
J.~Knechtel, ``{Resynthesis + SCOPE Setup},'' 2023,
  \url{https://github.com/DfX-NYUAD/resynthesis_attack}.

\bibitem{sisejkovic22}
D.~\v{S}i\v{s}ejkovi\'{c} \emph{et~al.}, ``Deceptive logic locking for hardware
  integrity protection against machine learning attacks,'' \emph{TCAD},
  vol.~41, no.~6, 2022.

\end{thebibliography}

\end{document}